\documentclass[conference,compsoc]{IEEEtran}
\IEEEoverridecommandlockouts

\usepackage{amssymb}
\usepackage{graphicx}
\usepackage{epstopdf}
\usepackage{gensymb}
\usepackage{color}
\usepackage{amsmath}
\usepackage{bm}
\usepackage{etoolbox}

\usepackage{cite}
\usepackage{array}
\usepackage{booktabs}
\usepackage{multirow}
\usepackage{comment}
\usepackage{subcaption}
\usepackage[ruled]{algorithm2e}
\usepackage{algorithmicx}
\usepackage{mathrsfs}
\usepackage{algpseudocode}

\usepackage{lastpage}
\usepackage{fancyhdr}

\usepackage{xcolor}
\usepackage[colorlinks,linkcolor={}]{hyperref}
\hypersetup{
  colorlinks,
  citecolor=black,
  linkcolor=black,
  urlcolor=blue}
\PassOptionsToPackage{hyphens}{url}\usepackage{hyperref}
\usepackage{enumitem}
\usepackage{wasysym}

\usepackage{balance}
\usepackage{soul}

\usepackage[framemethod=TikZ]{mdframed}
 
\alglanguage{pseudocode}

\algnewcommand{\Initialize}[1]{%
  \State \textbf{Initialize:}
  \Statex \hspace*{\algorithmicindent}\parbox[t]{.8\linewidth}{\raggedright #1}
}

\usepackage{threeparttable}
\usepackage{mathtools}
\usepackage{hyperref}
\usepackage{pifont}
\usepackage{listings}

\usepackage{url}
\urldef{\mailsa}\path|{alfred.hofmann, ursula.barth, ingrid.haas, frank.holzwarth,|
	\urldef{\mailsb}\path|anna.kramer, leonie.kunz, christine.reiss, nicole.sator,|
	\urldef{\mailsc}\path|erika.siebert-cole, peter.strasser, lncs}@springer.com|    
\newcommand{\keywords}[1]{\par\addvspace\baselineskip
	\noindent\keywordname\enspace\ignorespaces#1}
\usepackage[utf8]{inputenc}
\usepackage[english]{babel}

\usepackage{amsthm}
\usepackage{balance}

\usepackage{lipsum}
\usepackage{multicol}

\usepackage{bbding}

\theoremstyle{definition}

\definecolor{R}{RGB}{0,0,150}

\theoremstyle{remark}

\definecolor{myblue}{rgb}{0,0,0.9}

\definecolor{blue}{RGB}{60,132,196}
\definecolor{red}{RGB}{207,78,56}
\definecolor{gray}{RGB}{146,146,161}

\newcommand{\name}{DeepTheft\xspace}
\newcommand{\segment}{\texttt{$MetaModel_{Stru}$}\xspace}
\newcommand{\profile}{\texttt{$MetaModels_{Hyper}$}\xspace}
\newcommand{\eat}[1]{}

\newcommand{\mypara}[1]{\vspace{2pt}\noindent\textbf{{#1: }}}

\fancypagestyle{firstpage}{
  \fancyhf{}
  
  \fancyhead[C]{\vspace{10pt}\normalsize{To Appear in the 45th IEEE Symposium on Security and Privacy, May 20-23, 2024.}\\\vspace{-25pt}} 
  \fancyfoot[C]{\thepage}
}

\begin{document}

\title{\name: Stealing DNN Model Architectures through Power Side Channel}

\author{
   \IEEEauthorblockN{Yansong Gao\IEEEauthorrefmark{1}, Huming Qiu\IEEEauthorrefmark{2}, Zhi Zhang\IEEEauthorrefmark{3}, Binghui Wang\IEEEauthorrefmark{4}, \\ Hua Ma\IEEEauthorrefmark{5}, Alsharif Abuadbba\IEEEauthorrefmark{1}, Minhui Xue\IEEEauthorrefmark{1}, Anmin Fu\IEEEauthorrefmark{6}, Surya Nepal\IEEEauthorrefmark{1}}

   \IEEEauthorblockA{\IEEEauthorrefmark{1}CSIRO's Data61  \IEEEauthorrefmark{2}Fudan University \IEEEauthorrefmark{3}The University of Western Australia} 
   
   \IEEEauthorblockA{\IEEEauthorrefmark{4}Illinois Institute of Technology \IEEEauthorrefmark{6}Nanjing University of Science and Technology \IEEEauthorrefmark{5}The University of Adelaide}

   \thanks{Yansong Gao. Email: gao.yansong@hotmail.com}
   \thanks{Zhi Zhang is the corresponding author. Email: zhi.zhang@uwa.edu.au}
}

\maketitle

\thispagestyle{firstpage}

\begin{abstract}
Deep Neural Network (DNN) models are often deployed in resource-sharing clouds as Machine Learning as a Service (MLaaS) to provide inference services.
To steal model architectures that are of valuable intellectual properties, a class of attacks has been proposed via different side-channel leakage, posing a serious security challenge to MLaaS.

Also targeting MLaaS, we propose a new end-to-end attack, \name, to accurately recover complex DNN model architectures on general processors via the RAPL (Running Average Power Limit)-based power side channel. While unprivileged access to the RAPL has been disabled in bare-metal OSes, we observe that the RAPL is still legitimately accessible in a platform as a service, e.g., the latest docker environment of version 20.10.18 used in this work. 
However, an attacker can acquire only a low sampling rate (1\,KHz) of the time-series energy traces from the RAPL interface,
rendering existing techniques ineffective in stealing large and deep DNN models.
To this end, we design a novel and generic learning-based framework consisting of a set of meta-models, based on which \name is demonstrated to have high accuracy in recovering a large number (thousands) of models architectures from different model families including the deepest ResNet152. 
Particularly, \name has achieved a Levenshtein Distance Accuracy of 99.75\% in recovering network structures, and a weighted average F1 score of 99.60\% in recovering diverse layer-wise hyperparameters. 
Besides, our proposed learning framework is general to other time-series side-channel signals. To validate its generalization, another existing side channel is exploited, i.e., CPU frequency. Different from RAPL, CPU frequency is accessible to unprivileged users in bare-metal OSes. By using our generic learning framework trained against CPU frequency traces, \name has shown similarly high attack performance in stealing model architectures.

\end{abstract}

\section{Introduction}\label{sec:Intro}

\eat{
\begin{figure*}
	\centering
	\includegraphics[trim=0 0 0 0,clip,width=1 \textwidth]{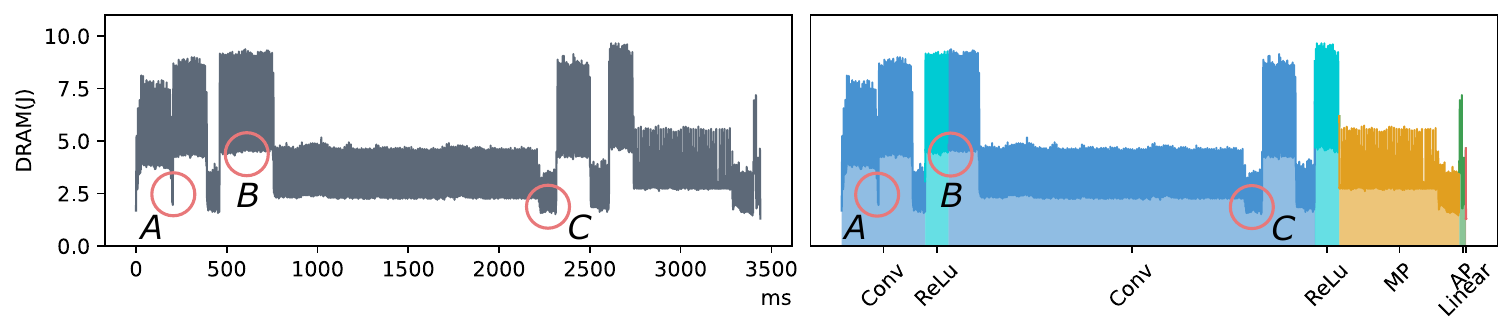}
	\vspace{-0.4cm}
	\caption{Highlighting {\bf Challenge 1}  by depicting a visualization of the DRAM channel for the power trace provided by RAPL. As shown by the red circle there is no reliable interlayer feature between layers. A and C are both sharp drop signals but not interlayer features, while B should have a sharp drop signal but is missing.}
	\label{fig:power}
	\vspace{-2mm}
\end{figure*}
}

Recent years have witnessed a huge success of Deep Neural Networks (DNN) in various areas, e.g.,  
pattern recognition, computer vision, and natural language processing. 
Generally, DNN models are often deployed in resource-sharing clouds to provide inference service, also called Machine Learning as a Service (MLaaS)~\cite{awsmlaas,googlemlaas}. 
A key factor determining a DNN's performance is its model architecture (i.e., comprised of DNN layer type and layer-wise hyperparameters), which is carefully designed via substantial human intelligence and intensive computing resources. Thus, DNNs' model architectures are regarded as valuable intellectual properties and high-value targets to an adversary~\cite{yan2020cache,wei2020leaky}. Additionally, a known DNN architecture can also be a premise to take further attacks such as model weights stealing~\cite{rakin2022deepsteal}, model hyperparameter stealing~\cite{wang2018stealing}, membership inference~\cite{shokri2017membership,carlini2022membership}, property inference~\cite{ganju2018property}, and data reconstruction~\cite{fredrikson2015model,balle2022reconstructing}. 

\mypara{Side-Channel DNN Model Architecture Stealing}
A number of works~\cite{naghibijouybari2018rendered,hua2018reverse,batina2019csi,yu2020deepem,xiang2020open,yan2020cache,wei2020leaky,zhang2021stealing,maia2021can,chmielewski2021reverse,wolf2021stealing,hu2020deepsniffer,zhu2021hermes,liang2022clairvoyance,meyers2022reverse} have been proposed to steal DNN model architectures based on different side-channel leakage (e.g., power~\cite{xiang2020open,chmielewski2021reverse,wolf2021stealing,zhang2021stealing,meyers2022reverse}, Electromagnetic (EM) emanations~\cite{batina2019csi,yu2020deepem,maia2021can,liang2022clairvoyance}, cache~\cite{yan2020cache}). 

These attacks can be classified into two categories. One category requires physical proximity or access to target machines hosting victim models~\cite{hua2018reverse,batina2019csi,yu2020deepem,xiang2020open,hu2020deepsniffer,maia2021can,zhu2021hermes,chmielewski2021reverse,wolf2021stealing,liang2022clairvoyance} as they need specialized equipment (e.g., magnetic sensors) to collect physical side-channel traces (e.g., magnetic signals) from target machines.
In this category, some attacks~\cite{hua2018reverse,batina2019csi} aim to steal shallow DNN models (fewer than 20 layers) running on edge devices (e.g., microprocessors) while others~\cite{hu2020deepsniffer,maia2021can} focus on attaining complex DNN models on advanced hardware (e.g., high-performance GPU). 

The other category of attacks avoids the above requirement,  where an attacker can co-reside with the same physical machine with victim models~\cite{wei2020leaky,yan2020cache}, thus posing a more significant threat to the MLaaS.
In this category, almost all attacks
target advanced hardware, as they are widely used in resource-sharing clouds.
In this MLaaS setting, 
Cache Telepathy~\cite{yan2020cache} is the only attack that infers a DNN model on general x86 CPUs via a cache side channel, as general processors are widely used by cloud platforms for model inference services~\cite{AWSCPU,hazelwood2018applied}. Please note that general CPUs are supported by cloud providers for model inference because their optimal cost-efficiency, e.g., AWS~\cite{AWSCPU}, Google~\cite{GoogleCPU} and Alibaba Clouds~\cite{AlibabaCPU} all supporting CPU inference.
In the same setting~\cite{yan2020cache}, we ask the following research question:
\begin{center}
  \emph{
    Is it feasible to leverage other side channel(s) on general CPUs to \emph{accurately} and \emph{stealthily} steal complex DNN architectures with legitimate system access?
  }
\end{center}

\mypara{\name} In this paper, we provide an affirmative answer to the question. We propose a new side-channel-based model architecture stealing attack, called \name, which targets x86 CPUs in the MLaaS. 
Orthogonal to the cache side channel~\cite{yan2020cache}, \name exploits RAPL (Running Average Power Limit)-based power side channel~\cite{lipp2021platypus,lipp2022amd}, for the first time, to perform the model architecture stealing. Note that \name is a stepping stone to weight stealing attacks either through side-channel~\cite{rakin2022deepsteal} or queried-based~\cite{oliynyk2022know}, which requires white-box knowledge of model architecture, especially for commonly used complex/deep models.

The RAPL is provided by Intel~\cite{lipp2021platypus} or AMD processors~\cite{lipp2022amd} as power management for software and has been widely used in clouds~\cite{sarood2013optimizing,zhang2016maximizing,guliani2019per}. 
Particularly, leveraging this power side channel enables an attacker to {passively} (i.e., being stealthy) collect energy-consumption traces when a DNN model is performing inferences.
We note that although access to the RAPL interface has been restricted to privileged software in the bare-metal OS setting, it is still accessible in the cloud setting. Particularly, we show that an attacker inside a container still has access to the RAPL interface and thus can read out the energy consumption of the system in the latest docker environment of 20.10.18, which is a prevalent containerization platform in practice~\cite{docker}, e.g., it has been commercialized by Amazon to deploy DNN models for model inference services~\cite{AWSdocker}. Also, we have experimentally confirmed that an attacker with a VM-privilege can extract power consumption through RAPL MSRs in the commercial cloud of AWS (e.g., an EC2 instance type of c4.8xlarge).

\name is motivated by~\cite{lipp2021platypus} that discovered the RAPL-based power side-channel on x86 processors, presenting a common characteristic with previous representative works (e.g., Cache Telepathy~\cite{yan2020cache} in USENIX SEC'20 and DeepSteal~\cite{rakin2022deepsteal} in Oakland'22), that is, all these works do not discover a side channel and they instead exploit an existing side channel to break the data confidentiality of deep learning. 
Particularly for \name, it is demonstrated to exploit the RAPL side channel to steal large and deep model architectures. 
To the best of our knowledge, the state-of-the-art (SOTA) side-channel attack~\cite{maia2021can} is the only existing one that steals full model architectures of 64 large and deep networks, which are prevalent in practice (e.g., ResNet101~\cite{he2016deep}).
While the SOTA requires physical proximity and a specialized probe to sample EM signals from a target high-end GPU at a rate of no less than 47\,KHz, it is also a learning-based approach.  We thus mainly compare \name with the SOTA~\cite{maia2021can}. 
For \name, the sampling rate of energy consumption from the RAPL user interface (i.e., \texttt{powercap}) is much lower, i.e., only 1\,KHz.

\mypara{Challenges} To this end, the constraint of \emph{low} sampling rate poses two technical challenges confronted with \name.
The first is to divide energy traces into segments of a target model's layers to accurately recover a complete network structure and enable subsequent layer-wise hyperparameters inferring. Due to the low sampling rate, 
we cannot observe salient boundaries among sampling points of different network layers, rendering existing techniques from the SOTA~\cite{maia2021can} ineffective in addressing this challenge. 
The second is to infer layer-wise hyperparameters. While some hyperparameters (e.g., kernel size) can be 
inferred with domain knowledge of model architectures~\cite{yan2020cache,maia2021can}, the rest cannot. 
{Particularly, important hyperparameters (i.e., kernel number for a convolutional layer and output feature size in a linear layer) have not yet been addressed by the SOTA, probably because both hyperparameters have a wide range of discrete integer values.}

\mypara{Our Solution} To address the challenges above, \name designs a novel learning framework that generalizes the extraction of large and deep models (e.g., ResNet152) on advanced hardware. 
Specifically, \name hybridizes i) U-Net~\cite{ronneberger2015u} that is widely used for image segmentation to capture spatial-feature, and ii) BiLSTM (Bidirectional Long Short-Term Memory~\cite{maia2021can}) to capture temporal-feature. Moreover, the performance of this hybrid model design is ensured by constructively-designed loss functions.
With this hybrid design, \name can recover all layers and divide an unknown energy trace into segments of sampling points that correspond to distinct layer types.
{Based on the segments of distinct layers, \name can then infer most layer-wise hyperparameters via common classification tasks. To infer the important hyperparameters of kernel number and output feature size, we note that there exists an inadvertent linear approximation between a segment of a convolutional (linear) layer and 
the computing overhead of the layer's operations. 
Further, the computing overhead is correlated with all the layer’s hyperparameters including kernel number (output feature size). 
With this insight, we perform a regression task against the computing overhead of convolutional (linear) layers, and then successfully infer kernel number (output feature size) based on the correlation. 
}

This newly designed learning framework is also applicable to other time-series side channel signals. For validation, \name exploits another recent revealed side channel, DVFS (dynamic voltage and frequency scaling)-enabled x86 CPU frequency side channel~\cite{wang2022hertzbleed}. Compared to the RAPL side channel, this frequency side channel also has a low sampling rate of 1\,KHz while it can be accessed unprivileged in the bare-metal OS setting. For the rest of the paper, we \textit{focus on} the RAPL side channel and present our evaluation of the frequency side channel in Section~\ref{sec:generic}.

Our contributions/results are summarized as follows\footnote{The source code and large-scale dataset are released at \url{https://github.com/LearningMaker/DeepTheft}.}:

\vspace{2pt}\noindent$\bullet$  
\name is an end-to-end attack, that, for the first time, demonstrates that either the power or CPU frequency side-channel can be used to successfully exfiltrate
large and deep DNN model architecture information on general processors.
In contrast to existing mainstream side-channel based model stealing attacks except~\cite{yan2020cache}, \name requires neither physical proximity nor access to target machines.

\vspace{2pt}\noindent$\bullet$ We propose a \textit{new general learning framework} that consists of a set of meta-models as existing work is ineffective in addressing newly raised challenges. Specifically,
In the offline phase, the framework learns the correlations between time-series signals (e.g., energy traces) and  DNN model architectures. In the online phase (where the stealing attack occurs), the pretrained meta-models surgically segment a time-series signal from an unknown model inference, resulting in an accurate network-structure recovery and timer-series signal segments. Further, the meta-models accurately infer layer-wise hyperparameters. We note that a single power trace per unknown model architecture is sufficient for stealing during the online inference phase.

\vspace{2pt}\noindent$\bullet$ We perform a comprehensive evaluation of \name  in an MLaaS setting of the latest docker environment.
Particularly, we construct a large training dataset including 50,000 energy traces
from 10,000 model architectures of prevalent model families to train the meta-models.  
In the online phase, the pretrained meta-models are tested against a test dataset of 5,760 energy traces from 1,152 unknown model architectures. The experimental results show that \name 
demonstrates a high LDA of 99.75\% in recovering network structures, and a high weighted average F1 score of 99.60\% in inferring diverse layer-wise hyperparameters,  including the kernel number and the output-feature size that have not been addressed by the SOTA~\cite{maia2021can}.
Also, we compare \name with the SOTA by using their open-source code in the same setting, results of which show that \name performs much better (41\% LDA higher) than the SOTA in recovering the network structure. 

\vspace{2pt}\noindent$\bullet$ We further validate the generalization of the learning framework by using another time-series side channel signal, i.e., CPU-frequency, in a bare-metal Ubuntu 18.04.6 LTS server. 
Our evaluation shows that \name with the CPU-frequency side channel has achieved a high attacker performance in stealing model architectures, similar to that with the power side channel.

\vspace{0.10cm}
\noindent{\bf Responsible Disclosure.} \name exploits a publicly known power side channel\footnote{See CVE-2020-8694 and CVE-2020-8695 from \url{https://platypusattack.com/}.} and a CPU frequency side channel\footnote{See CVE-2022-23823 (AMD) and CVE-2022-24436 (Intel) from \url{https://www.hertzbleed.com/}} and thus there is no need to report it to the CPU vendors. This is similar to DeepSteal (Oakland'22)~\cite{rakin2022deepsteal} that exploited a publicly known rowhammer-based side channel\footnote{See CVE-2019-0174 from \url{https://rambleed.com/}.} to steal model weights. However, as RAPL is still accessible to docker containers, we have disclosed  this security issue to the docker security team on 13th, July, 2023. The team has acknowledged it and set an embargo date until 19th, September, 2023 before releasing its fix.

\section{Background}
\label{sec:bkgd}

\subsection{Deep Neural Network (DNN)}
A DNN is composed of multiple layers of interconnected artificial neurons to map the linear or non-linear relationship between input data and output data. A DNN has two stages: training and inference.  
As the training process is computationally heavy, it is often performed on GPUs. However, attributing to cost-efficiency, CPUs are also used to support the DNN inference for good responsiveness~\cite{yan2020cache} e.g., AWS~\cite{AWSCPU}, Google~\cite{GoogleCPU} and Alibaba Clouds~\cite{AlibabaCPU} all supporting CPU inference.

\mypara{DNN Architecture}
As substantial human intelligence and intensive computing resources are often required to develop DNN model architectures, the architecture has become the intellectual property and a primary target for attackers. A DNN architecture is decided by a list of \emph{hyperparameters} including the structure and layer-wise hyperparameters. Aligned with previous model-stealing attacks~\cite{wei2020leaky,yan2020cache,maia2021can}, 
we aim at stealing the following hyperparameters:

\vspace{2pt}\noindent$\bullet$ the number of layers (e.g., ResNet152 has 152 layers);

\vspace{2pt}\noindent$\bullet$ the type of each layer (e.g., fully-connected layer, convolutional layer and pooling layer);

\vspace{2pt}\noindent$\bullet$ the activation function for each layer (e.g., \texttt{relu});

\vspace{2pt}\noindent$\bullet$ the layer-wise hyperparameters for a specific layer, e.g., the neuron number for a fully-connected layer; the kernel size for a convolutional layer, the stride size for a pooling layer.

\subsection{RAPL}
Intel has introduced the RAPL feature to enable power management since its SandyBridge microarchitecture~\cite{zhang2021red,lipp2021platypus}. 
Particularly, the Intel RAPL provides energy consumption of underlying hardware to software at a granularity of the so-called \emph{domain}. 
Intel defines five main domains, that is, \emph{Package} is the entire CPU socket; \emph{Power planes 0 and 1} (PP0 and PP1) refer to the processor cores and uncore devices on the socket, respectively; \emph{DRAM} is the main memory attached to the memory controller; \emph{Platform} covers the whole system-on-chip. 

Moreover, multiple model-specific registers (MSRs) are defined to serve the energy estimation and they can be accessed by privileged software (e.g., kernel).
To help unprivileged software acquire energy estimates for each domain, Intel has implemented a power capping framework for Linux (i.e., \texttt{powercap}) since Linux kernel 3.13, which reads values via relevant MSRs and provides them to user-space access via the interface named \texttt{sysfs}. Since the RAPL side-channel has been discovered~\cite{lipp2021platypus}, only root-privilege users have meaningful access to the \texttt{sysfs} interface, that is, read access by unprivileged users returns 0 only. 

Nevertheless, at the time of writing this paper, we observe that the interface is still accessible to a container in the latest docker version of 20.10.18. Additionally, certain AWS EC2 instances are allowed to access RAPL MSRs. 
While \name is demonstrated on the Intel CPUs, we note that AMD CPUs also introduce the RAPL feature since its Zen microarchitecture and even provide energy estimates at the per-core level, a finer granularity compared to the PP0 domain in the Intel CPUs.

\section{Related Work}\label{sec:related}

\mypara{Side-Channel based Model Architecture Stealing} A large number of attacks~\cite{naghibijouybari2018rendered,wei2018know,hua2018reverse,batina2019csi,yu2020deepem,xiang2020open,yan2020cache,wei2020leaky,zhang2021stealing,tian2021remote,maia2021can,chmielewski2021reverse,wolf2021stealing,hu2020deepsniffer,zhu2021hermes,liang2022clairvoyance,rakin2022deepsteal,meyers2022reverse} have been proposed to leverage different side-channel leakage to break data confidentiality relevant to DNN security, among which some representative works and \name are shown in \autoref{tab:attacks}. While a few works can extract model parameters such as weights~\cite{rakin2022deepsteal}\footnote{In~\cite{rakin2022deepsteal}, an attacker is assumed to know the model architecture of the victim model as prior knowledge.} or recover model inputs~\cite{wei2018know,tian2021remote}, most of the them~\cite{naghibijouybari2018rendered,hua2018reverse,batina2019csi,yu2020deepem,xiang2020open,yan2020cache,wei2020leaky,zhang2021stealing,maia2021can,chmielewski2021reverse,wolf2021stealing,hu2020deepsniffer,zhu2021hermes,liang2022clairvoyance,meyers2022reverse} focus on stealing model architectures. 
Existing model-stealing attacks generally assume no prior knowledge about a target model and rely on different side-channel leakage sources, e.g., power~\cite{xiang2020open,chmielewski2021reverse,wolf2021stealing,zhang2021stealing,meyers2022reverse}, memory-access pattern~\cite{hua2018reverse}, computer bus~\cite{hu2020deepsniffer,zhu2021hermes}, electro-magnetic (EM) emanations~\cite{batina2019csi,yu2020deepem,maia2021can,liang2022clairvoyance}, cache~\cite{yan2020cache}, or GPU features (i.e., context switch~\cite{wei2020leaky} and performance counters~\cite{naghibijouybari2018rendered}).
In the following, we discuss these attacks based on whether they require physical proximity or access to a target machine where a victim model is residing. The \name does not require such proximity/physical access.

\begin{table}[h]
\vspace{+4mm}
	\centering 
	\caption{\name and representative works (in chronological order) that use side-channels to compromise data confidentiality of deep learning. }
	\resizebox{0.50\textwidth}{!}{
	\begin{tabular}{l | c c c c }  
		\toprule 
		 {\bf Work} & {\bf Side Channel} & {\begin{tabular}[c]{@{}c@{}}{\bf No Physical}\\ {\bf Proximity/Access?}\end{tabular}}  &  {\begin{tabular}[c]{@{}c@{}}{\bf Targeted}\\ {\bf Hardware}\end{tabular}}\\ 
		\midrule

		 {\begin{tabular}[c]{@{}c@{}} Hua \textit{et al.}~\cite{hua2018reverse}  (DAC'18) \end{tabular}}
		 & Memory Access Pattern  & ${\times}$ & FPGA \\ \midrule
         {\begin{tabular}[c]{@{}c@{}} CSI NN~\cite{batina2019csi} (USENIX SEC'19) \end{tabular}} & Electro Magnetic Emanation & ${\times}$  & MCU \\ \midrule
        {\begin{tabular}[c]{@{}c@{}} DeepSniffer~\cite{hu2020deepsniffer} (ASPLOS'20) \end{tabular}} & Computer Bus  & ${\times}$ & GPU \\ \midrule
        {\begin{tabular}[c]{@{}c@{}} Cache Telepathy~\cite{yan2020cache} (USENIX SEC'20) \end{tabular}}
		 & Cache & ${\surd}$ & CPU \\ \midrule

        {\begin{tabular}[c]{@{}c@{}} Hermes Attack~\cite{zhu2021hermes} (USENIX SEC'21) \end{tabular}} & Computer Bus & ${\times}$ & GPU \\ \midrule

		 {\begin{tabular}[c]{@{}c@{}} DeepSteal~\cite{rakin2022deepsteal} (Oakland'22) \end{tabular}}
		 & Rowhammer & ${\surd}$ & DRAM \\ \midrule
		 
		 {\begin{tabular}[c]{@{}c@{}} Maia \textit{et al.}~\cite{maia2021can} (USENIX SEC'22) \end{tabular}}
		 & Electro Magnetic Emanation & ${\times}$  & GPU \\ \midrule
		{\begin{tabular}[c]{@{}c@{}}\textbf{\name} (Ours) \end{tabular}}
		 & Power (and CPU Frequency) & ${\surd}$ & CPU\\

		\bottomrule
	\end{tabular}
			}
	\label{tab:attacks} 
        \vspace{-1mm}
\end{table}

\noindent$\bullet$ \textit{With Proximity/Physical Access.} To recover a victim model architecture, one class of side-channel attacks relies on customized equipment for side-channel information collection and requires either physical proximity~\cite{batina2019csi,yu2020deepem,xiang2020open,maia2021can,chmielewski2021reverse,wolf2021stealing,liang2022clairvoyance} or physical access~\cite{hua2018reverse,hu2020deepsniffer,zhu2021hermes} to targeted machines. 
Batina \textit{et al.}~\cite{batina2019csi} leveraged both model querying and EM probing to extract shallow neural networks (no more than 20 layers) running on edge microcontrollers (MCUs).
Yu \textit{et al.}~\cite{yu2020deepem} also leveraged the EM side channel to recover binarized neural networks that are tailored for edge devices.
Chmielewski \textit{et al.}~\cite{chmielewski2021reverse} performed a simple EM and power analysis to partially recover targeted MLP and CNN models (e.g., CNN lay-wise hyperparameters are not yet extracted) on an edge device (i.e., Nvidia Jetson Nano). 
Wolf \textit{et al.}~\cite{wolf2021stealing} conducted a simple power analysis using a specialized device (i.e., a ChipWhisperer Lite) against an MCU to differentiate one model from another. 
Xiang \textit{et al.}~\cite{xiang2020open} revealed model architectures via power traces in a raspberry-pi platform. 
Different from prior works that exploited near-field EM signals, Liang \textit{et al.}~\cite{liang2022clairvoyance} collected far-field EM traces via a specialized device, which can be several meters away from a victim machine.
Their proposed attack was demonstrated on an NVIDIA GPU to recover shallow model architectures. 
Hua \textit{et al.}~\cite{hua2018reverse} analyzed off-chip memory-access patterns via a hardware trojan insertion and reverse-engineered structure and weights of a shallow model running at an edge FPGA accelerator. 
Hu \textit{et al.}~\cite{hu2020deepsniffer} combined EM traces and bus snooping to generate statistics of memory reads/writes, patterns of which are analyzed to infer complex model architectures on advanced GPU. Different from them, Maia \textit{et al.}~\cite{maia2021can} could recover layer-wise hyperparameters for extracted layers more precisely by analyzing GPU-emitted magnetic signals via a physical sensor. 
Similar to ~\cite{hu2020deepsniffer}, Zhu \textit{et al.}~\cite{zhu2021hermes} developed a bus snooping approach to capture the PCIe traffic between host machines and advanced GPU devices and reconstruct the target model architectures. 

\noindent$\bullet$ \textit{Without Proximity/Physical Access.} Compared to the aforementioned class of attacks, the other class of side-channel attacks~\cite{naghibijouybari2018rendered,yan2020cache,wei2020leaky,zhang2021stealing,meyers2022reverse} that need co-location between an attacker and a victim in the same physical machine are more applicable to the MLaaS clouds where physical proximity or physical access is almost impossible to an attacker.
With an in-depth code analysis of the implementation of Generalized Matrix Multiply, Yan \textit{et al.}~\cite{yan2020cache} leveraged a cache side channel in x86 CPU to infer the architecture of a complex model. 
Naghibijouybari \textit{et al.}~\cite{naghibijouybari2018rendered} abused the CUDA profiling tools interface of Nvidia GPU as performance counters to sample performance events from a victim DNN model and thus partially infer the model architecture, i.e., the number of neurons of the model's input layer. 
To address the limitation, Wei \textit{et al.}~\cite{wei2020leaky} exploited the penalty of Nvidia GPU's context switches. Their proposed attack could slow down the transition between DNN layer operations to collect enough samples per DNN layer operation and improve its accuracy for predicting a shallow model architecture. 
Zhang \textit{et al.}~\cite{zhang2021stealing} used a power sensor (e.g., ring oscillators) to collect power traces from a victim model in a shared FPGA platform that supports multi-tenancy. 
Also in this shared FPGA setting, Meyers \textit{et al.}~\cite{meyers2022reverse} took network folding into consideration and recovered the folding parameters with power side-channel analysis.
We note that the multi-tenancy FPGA platform has not yet been commercialized whereas the current cloud provides only support for single-tenancy FPGA.

\mypara{Query based Model Stealing} 
Query based attack mainly attempts to obtain a model that shares similar behavior to the victim model (i.e., prediction between the stolen model and victim model is consistent, namely model fidelity) and cannot steal model architectures~\cite{carlini2020cryptanalytic}. In this attack, the attacker queries the victim model with (limited) number of samples and gains corresponding outputs (e.g., \texttt{softmax} or single label) 
provided by the victim model. Those sample-label pairs are then used to train a substitute model that attempts to have comparable fidelity to the victim model. Therefore, the attacker can avoid paying for queries later to the model provider. 
However, sample query based model stealing~\cite{tramer2016stealing,wang2018stealing} is inefficient against complicated models unless the model architecture is known~\cite{oliynyk2022know}.

\begin{figure*}[!t]
	\centering
	\includegraphics[trim=0 0 0 0,clip,width=0.80\textwidth]{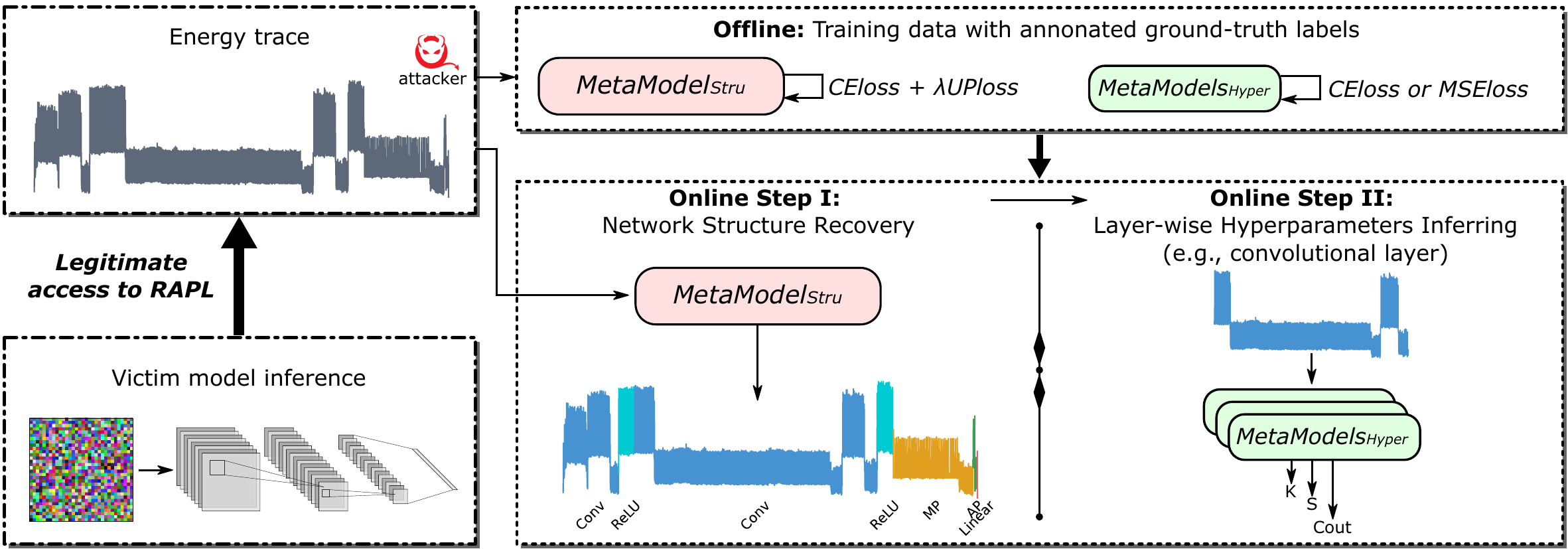}
	\caption{An overview of \name.}
	\label{fig:overview}
	\vspace{-5mm}
\end{figure*}

\section{\name}\label{sec:implementation} 

In this section, we first give our threat model, and then introduce the overview and design details of \name.
 
\subsection{Threat model}\label{sec:threatmodel} 
Aligned with previous side-channel based model stealing attacks~\cite{batina2019csi,yu2020deepem,yan2020cache,rakin2022deepsteal,maia2021can}, we assume an attacker aims to steal the architecture (i.e., network structure and layer-wise hyperparameters) of a target DNN model providing the inference service. We further assume the attacker can leverage prior techniques~\cite{ristenpart2009hey,zhang2014cross,varadarajan2015placement,ezhilchelvan2015evaluating,atya2017malicious} to co-locate herself onto the targeted machine as the victim model, which is practical in a shared-cloud scenario, e.g., the prevalent MLaaS~\cite{ribeiro2015mlaas}. 
In our implementation, we use the recent docker environment of version 20.10.18 to simulate a container cloud for acomprehensive evaluation~\cite{gao2017containerleaks}.
Particularly, the attacker, residing in a docker container, has legitimate access to the \texttt{powercap} interface but is not allowed to access another docker container running the model, as the docker environment in the targeted machine is benign and deploys proper container isolation. All the container images can be different and they are pulled from the official Docker hub. In \autoref{sec:dis}, we also demonstrated the potential of using \name for model stealing in the AWS setting where an attacker residing in an EC2 instance is allowed to extract the energy consumption via RAPL MSRs.
Last, \name is a black-box attack assuming no prior knowledge about the targeted DNN model.
The attacker can only feed inputs to the target model through a legitimate query interface~\cite{tramer2016stealing}, passively and non-intrusively collect the energy traces from the shared machine and analyze them 
for model-architecture stealing. {Given an unknown model architecture, a single query or snapshot fed into the unknown running model is sufficient in stealing the model}.

\subsection{Overview}\label{sec:overview}

\name has two phases, i.e., offline and online, as shown in \autoref{fig:overview}. In its offline phase, \name develops a constructive learning framework that consists of a set of meta-models.
These models are trained upon a dataset, which has a large number of energy traces from large neural networks of various model architecture families (e.g., AlexNet, VGG, ResNet, and random architectures). For each energy trace, the attacker performs sampling for all available domains via the \texttt{powercap} interface to collect sampling points. 
The sampling can be conducted from a machine model same to the victim machine. After that, each sampling point is labeled with a ground-truth layer type (e.g., linear layer, convolutional layer and pooling layer). 

In its online phase, \name queries a black-box target model in the victim machine to trigger its model inference and capture its energy trace. With the energy trace, \name implements a two-step attack. In the first-step, a pre-trained meta-model gained in the offline phase, coined \segment, is used to perform a surgical segmentation against the energy trace, resulting in segmented points corresponding to different layers of an accurate network structure (e.g., the number of layers and the layer type). In the second step, with the segmented points of each layer, \name leverages multiple other pretrained meta-models (called \profile) and domain knowledge to infer layer-wise hyperparameters.

Clearly, the performance of the meta-model training in the offline phase is critical to the effectiveness of the model architecture stealing in the online phase. In the online phase, the accuracy of \segment affects the accuracy of \profile, as inferring the layer-wise hyperparameters 
is dependent on the trace segmentation that segments the trace into layers precisely---all points corresponding to a layer are fed into \profile for layer-wise hyperparameters inferring.
To this end, the challenges of \name are to acquire a well-trained \segment and \profile. 

Specifically, the first challenge is to train \segment so that it can divide an unknown energy trace into segments of sampling points (each segment predicted with a layer type), and enable subsequent layer-wise hyperparameters inferring for \profile. 
For this challenge, an intuitive solution is to leverage Bidirectional Long Short-Term Memory (BiLSTM)~\cite{graves2005bidirectional} from the SOTA~\cite{maia2021can} for capturing the temporal features (e.g., the chronological sequence of the sampling points) of a given energy trace.
However, the SOTA has a high sampling rate (no less than 47\,KHz) 
and thus can observe the salient boundaries among the sampling points caused by the execution of different layers. For \segment, applying BiLSTM would result in points mis-segmentation, that is, some points will be mis-classified into a wrong layer type (we compared \segment with BiLSTM in the SOTA in~\autoref{sec:compressionComp}), as \name has a sampling rate of only 1\,KHz.
To address this challenge, \segment is inspired by U-Net~\cite{ronneberger2015u}, a popular model architecture for extracting spatial features of 2D (Dimensional) images.
In \segment, the U-Net is used to extract the spatial features of the 1D energy traces, e.g., the sampling points in a trace can exhibit distinct visual shapes. However, the U-Net itself is ineffective to capture temporal features provided by the time-series energy trace and missing the rich temporal information per se. \segment addresses this by incorporating BiLSTM to capture both spatial features with the U-Net, which form a new hybrid architecture.
Further, a new loss function is carefully designed for \segment to learn correlations among sampling points from within-one-layer to across-layers.

The second challenge is to train \profile so that it can infer layer-wise hyperparameters for a given segment of sampling points (each segment is marked by \segment with a layer type). 
While some hyperparameters can be decided using domain knowledge (e.g., the input kernel number is equal to the output kernel number of a convolutional layer), there are 7 hyperparameters (see \autoref{tab:hyperparameters}) that need to be inferred using \profile.
{Among them, 5 hyperparameters (e.g., kernel size in a convolutional layer) have a much small range of value candidates and thus \profile can infer them via common classification tasks. 
The challenge is to infer the kernel number in a convolutional layer and the output feature size in a linear layer, as both have a wide range of discrete integer values. (e.g., $2^n$ where $n$ can be from $\{4,5,6, \cdots\}$) 
and they can hardly be covered by the energy traces. Thus, {it is hard to infer them via a classification task.} 
It is also inaccurate to perform regression tasks against them as there is no direct correlation between the energy traces and either of them. 
This might be the reason why the SOTA~\cite{maia2021can} did not infer them. 
To address this challenge, we observe that for a segment of sampling points for a specific layer-type, it has linear approximation with the computing overhead of the layer's operations, which are correlated by all the layer's hyperparameters. 
Thus, either the kernel number or the output feature size is one of the hyperparameters that correlate linearly with the computing overhead. With this observation, we perform an indirect regression against the computing overhead of convolutional/linear layers, and then indirectly infer the hyperparameter (kernel number/output feature size) based on its correlation with the overhead.}
To this end, we customize \segment to generate \profile's model architectures for either classification tasks or indirect regression tasks.  

In the following, we discuss the design of \segment in \autoref{sec:seg}, domain knowledge in \autoref{sec:domain} and \profile in \autoref{sec:hymetamodel}.

\begin{figure*}[t]
	\centering
	\includegraphics[trim=0 0 0 0,clip,width=0.80 \textwidth]{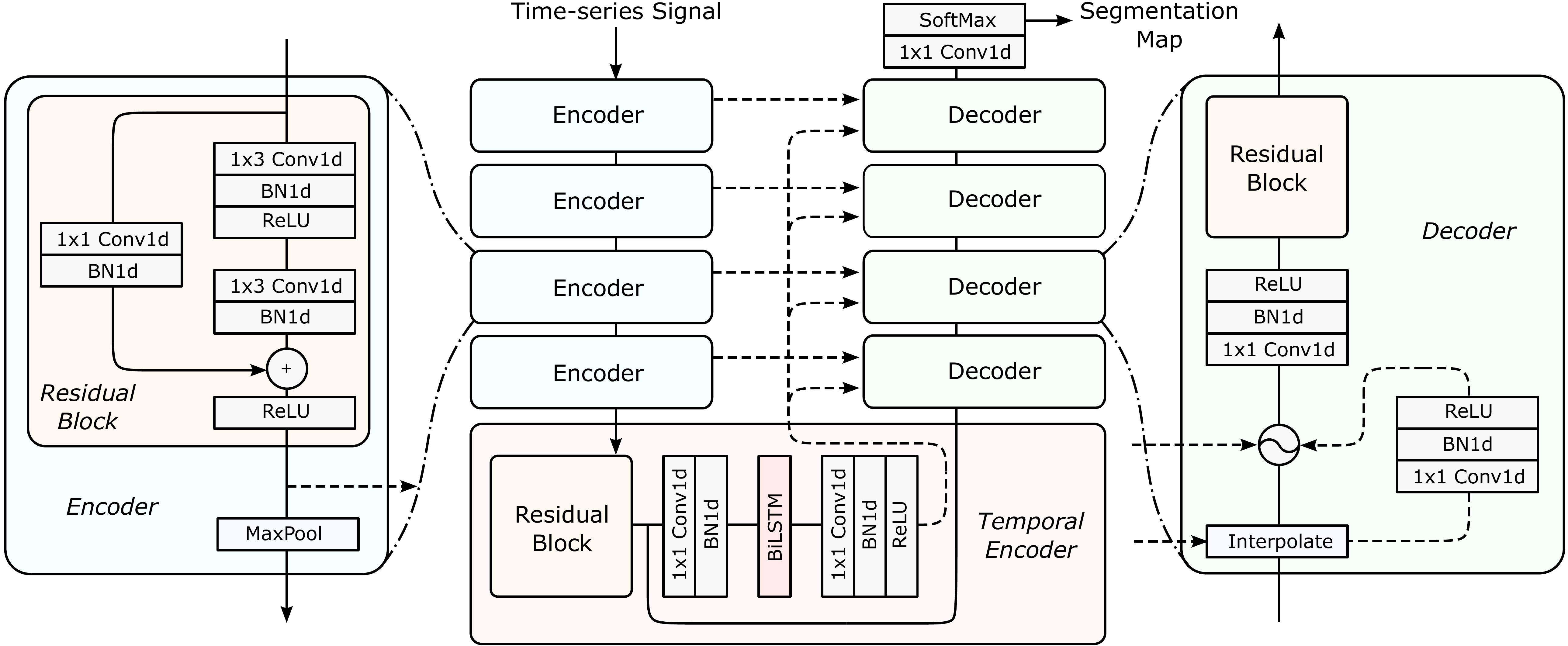}
	\caption{The model architecture of \segment. Its input is a time-series signal (e.g., an energy trace of sampling points for \name). Its output is a segmentation map, which is a matrix of predicted probabilities of layer types for all sampling points in the trace.
	}
	\label{fig:detailmodel}
	\vspace{-6mm}
\end{figure*}

\subsection{\segment Design}\label{sec:seg}
We first introduce the model architecture of \segment, and then discuss how to formulate its loss function for effective learning. 
\subsubsection{Model Architecture}

As shown in \autoref{fig:detailmodel}, the backbone network of \segment is a hybrid design of U-Net and BiLSTM, which extracts spatial features and temporal features from an input time-series energy trace, and outputs a segmentation map, which is a matrix of predicted probabilities of layer types for all sampling points in the trace. 

\mypara{Capturing Spatial Features via U-Net} 
U-Net is made up of a set of 1D CNNs, which consists of a down-sampling path (contracting path) on the left of \autoref{fig:detailmodel},  and an up-sampling path (expanding path) on the right of the figure.
The down-sampling path has four encoders, which down-sample a spatial feature map step-by-step using the maxpooling operation and thus gradually aggregate the spatial information into semantic information. 
In this process, the spatial feature map is eventually abstracted into a rich-semantics feature map. 
The up-sampling path has four symmetric decoders, which up-sample the abstracted feature map step-by-step using linear interpolation and thus gradually restore the spatial information with the semantic information preserved. 

Both the encoders and decoders are based on residual blocks to mitigate the gradient vanish issue~\cite{he2016deep}. To compensate for the information lost from the 
down-sampling path, a feature map output by each encoder is fused via a channel (i.e., a skip-connection operator denoted as a dashed line with an arrow in \autoref{fig:detailmodel}) into a decoder along the up-sampling path. By fusing the low-level features with high-level features, the decoders can restore the encoders' abstracted information from the time-series trace, thus improving \segment's performance.

\mypara{Capturing Temporal Features via BiLSTM} 

At the bottom of \autoref{fig:detailmodel}, a BiLSTM-based {temporal encoder} is placed, which takes the spatial feature map from the last encoder of U-Net as input
and enhances the learning of temporal features. 
As the input of the temporal encoder is the smallest spatial feature map (downsampled by 16 times) from U-Net, it 
improves the BiLSTM's training efficiency. Besides, its high-level semantics facilitate the capture of temporal features. Thus, the output of the temporal encoder is {rich in temporal features}. To better combine temporal features and spatial features, the temporal encoder provides an additional channel (a skip-connection operator) as an input to each decoder. 

As the input to \segment is an
energy trace from a model inference, its length can be varied. Thus, we pad the input with a factor of 16, given that the 4 encoders downsample the energy trace by a factor of 16. We note that \segment is designed to generalize the time-series energy trace segmentation and thus it is also applicable to other time-series signals, such as the EM signals from the SOTA~\cite{maia2021can}.

\subsubsection{Loss Function in Training}\label{sec:lossfunc}
Suppose we have obtained a dataset consisting of $N$ energy traces $\mathcal{D}=\{\textbf{x}^i,\textbf{y}^i,\textbf{z}^i\}_{i=1}^N$ (the dataset construction is deferred to ~\autoref{sec:datasetCon}). Here, each $\textbf{x}^i \in \mathbb{R}^{c \times l}$ is an energy trace of performing a model inference. Each trace has $c$ channels corresponding to the RAPL's domains (e.g., \texttt{DRAM}) and $l$ samples (e.g., one sampling point every $1\,ms$ in our dataset as the sampling rate is 1\,KHz), with each sample $\textbf{x}^i_{s}$ having a {layer type} $\textbf{y}^i_{s}$, a $k$-dim  one-hot encoding vector 
for $k$ layer types (e.g., pooling layer, convolutional layer, ReLU). 
Thus, the aforementioned segmentation map is a $l\times k$ matrix.
$\textbf{z}^i \in \mathbb{N}^{n \times 2}$ is the position label matrix for $n$ layers in the energy trace $\textbf{x}^i$, where each row records the start and end position of the respective layer. We note that $\textbf{z}^i$ is only required during the offline phase to train the meta-models and will not be used in the online phase.
With this dataset, we discuss how to design our loss function for \segment.

Specifically, the model is denoted as $f$ with parameters $\theta$, and 
$f_\theta(\textbf{x}^i) \in \mathbb{R}^{k \times l}$ is to map an energy trace  $\textbf{x}^i \in \mathbb{R}^{c \times l}$ to its confidence scores, where $f_\theta(\textbf{x}^i)_{s}$ indicates the probability vector of predicting a sample point $s$ in ${\bf x}^{i}$.

\mypara{Sampling-Point-Independent Loss} 
A possible way to train \segment is by 
minimizing the standard cross-entropy (CE) loss as below: 
\vspace{-2mm}
\begin{align}
\label{eq_2}
\mathcal{L}_{CE} = 
-\sum_{{\bf x}^i,{\bf y}^i \in \mathcal{D}} \sum^{l}_{s} 
{\bf y}^i_{s} \log(f_\theta({\bf x}^{i})_s).
\end{align}
\vspace{0mm}
However, training the model with the CE loss alone cannot achieve high accuracy for two main reasons.  
First, the CE loss is the sum of the losses of all sampling points with the same weight (i.e., 1) and the number of points corresponding to different layers can be highly imbalanced. Take an energy trace we collected as an example, one convolutional layer occupies more than half of the points, while a linear layer has as few as 1 or 2 points. Thus, the CE loss is not effective in handling imbalanced sampling points, as minimizing it would  
mislead the model to predict all sampling points into the layer types with more points, making the layer type with few points ignored.
Second, considering that the CE loss predicts a layer type per sampling point, it 
cannot capture the contextual information among neighbors of a point and thus cannot ensure that the predicted layer types for the neighbors are consistent.

\mypara{Cross-Sampling-Point Loss} 

To address the aforementioned problem, we propose a complementary loss, termed {Unique Path (UP) Loss}, which is inspired by CTCLoss~\cite{graves2006connectionist}, a solution mainly used to segment non-sequential data. Compared to the CTCLoss, the UP loss has a different design for two considerations.
First, the CTCLoss allows identical layer prediction in its prediction sequence, which does not hold for \segment as a model does not have consecutive identical layers. 
Second, as {inputs are unnecessarily aligned with outputs}, the CTCLoss needs to find all possible {paths} to make a correct prediction, which incurs a high computational overhead.  
For \segment, {its inputs and outputs are inherently aligned}, and thus the UP loss only needs to identify the unique possible path 
for a correct prediction, which is much more efficient.

\begin{figure}[h]
	\centering
	\includegraphics[trim=0 0 0 0,clip,width=0.48 \textwidth]{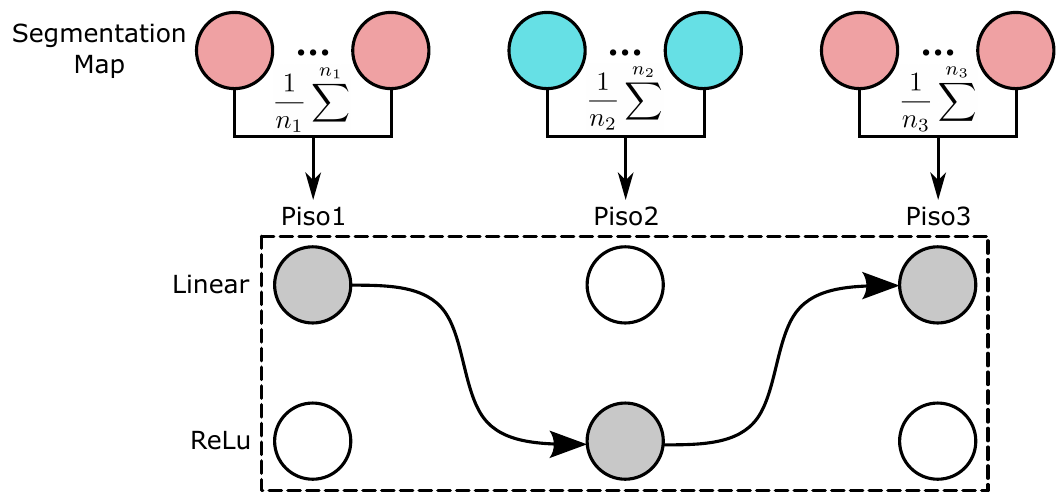}
	\vspace{+2mm}
	\caption{An illustrative example of using UPL in an MLP with three layers [Linear, ReLu, Linear]. It first averages the probabilities of the successive classes in the segmented map and aggregates them to the  probabilities ($P_{iso}$) of the isolated sampling points, and then  multiplies $P_{iso}$ to all layers to obtain the unique path probability.}
	\label{fig:UPL}
	\vspace{-1mm}
\end{figure}

Specifically, we first apply each position label ${\bf z}^i_m$ to isolate the sampling points in ${\bf x}^i$ corresponding to a layer. Then we compute the {unique path probability} for this layer, which is defined as the average prediction probability of the isolated sampling points in this layer as:
\begin{equation}\label{eq_3}
P_{iso}({\bf z}^i_m) = 
\frac{
\sum_{s={z}^i_{m,0}}^{{z}^i_{m,1}}
\textbf{y}^i_{s} f_\theta({\bf x}^i)_{s}
}
{{z}^i_{m,1}-{z}^i_{m,0}+1}.
\end{equation}
In principle, the layer corresponding to $P_{iso}({\bf z}^i_m)$ is deterministic in a one-to-one manner, and  $P_{iso}({\bf z}^i_m)$ can represent the probability of all sampling points within ${\bf z}^i_m$ being predicted correctly.
Then, we multiply $P_{iso}({\bf z}^i_m)$ for all layers in the energy trace to represent the unique path probability for all layers as shown in \autoref{fig:UPL}, which can measure the closeness between all the predicted layers and the ground-truth layers. 
Formally, the UPL is defined 
as follows: 
\begin{equation}\label{eq_4}
\mathcal{L}_{UP} = 
-\sum_{{\bf z}^i \in \mathcal{D}}
\log(\prod_{m=1}^n P_{iso}({\bf z}^i_m)).
\end{equation}
The UPL captures the relationships among neighboring points and minimizing it encourages consistent prediction of sampling points from the same layer.
As the UPL is defined on layers instead of sampling points, it can also alleviate the issue of imbalanced sampling points of different layers.

\mypara{Loss Function Formulation for \segment} The ultimate loss used to train the \segment is formulated as below:
\begin{equation}\label{eq_5}
\mathcal{L} = 
\mathcal{L}_{CE} + \lambda \mathcal{L}_{UP}, 
\end{equation}
where $\lambda$ is a hyperparameter to balance the two losses (e.g., $\lambda = 1$ by default). 
To this end, $\mathcal{L}_{CE}$ aims to achieve an accurate prediction per sampling point, and  $\mathcal{L}_{UP}$ is carefully designed to learn the context information of sampling points from within a layer (i.e., isolating points from a layer via ${\bf z}^i_m$) and different layers (i.e., maximizing the unique path probability for all layers). Hence, the UP loss 
can capture both spatial and temporal features. 

\begin{table}
\vspace{+4mm}
\centering 
\caption{A list of layer-wise hyperparameters for the three layers. 
If those marked in {\color{red} red} are inferred via \profile, the rest can then be decided (Hypers, Conv and MP denote Hyperparameter, Convolutional, and Max Pooling, respectively).} 
\resizebox{0.4 \textwidth}{!}
{
\begin{tabular}{c | ll  }
\toprule
{\bf Layer} & {\bf Hypers} & {\bf Description} \\ 

\midrule
\multirow{10}*{Conv}
& $C_{\rm in}$ & \begin{tabular}{@{}l@{}}Number of channels of an input. \\Its value is decided by the input or $C_{out}$.\end{tabular} \\
\cline{2-3}
& {\color{red} $C_{\rm out}$} & \begin{tabular}{@{}l@{}}Number of kernels. \\Its value is $2^n$ where $n$ can be from $\{4,5,6, \cdots\}$.\end{tabular}\\
\cline{2-3}
& {\color{red} $K$} & \begin{tabular}{@{}l@{}}Size of a kernel. Its value is from $\{1, 3, 5, 7\}$.\end{tabular}\\
\cline{2-3}
& {\color{red} $S$} & \begin{tabular}{@{}l@{}}Stride. Its values is from $\{1, 2\}$.\end{tabular}\\
\cline{2-3}
& $P$ & \begin{tabular}{@{}l@{}}Size of padding. \\Its value is decided by $\lfloor (K - 1) / 2 \rfloor \times D$.\end{tabular}\\
\cline{2-3}
& $D$ & \begin{tabular}{@{}l@{}}Dilation. Its value is set to 1.\end{tabular}\\
\cline{2-3}
& $G$ & \begin{tabular}{@{}l@{}}Groups. Its value is set to 1.\end{tabular}\\

\midrule
\multirow{4}*{MP}
& {\color{red} $K$} & \begin{tabular}{@{}l@{}}Size of a kernel. Its value is from $\{2, 3\}$.\end{tabular}\\
\cline{2-3}
& {\color{red} $S$} & \begin{tabular}{@{}l@{}}Stride. Its value is from $\{1, 2\}$.\end{tabular}\\
\cline{2-3}
& {\color{red} $P$} & \begin{tabular}{@{}l@{}}Size of padding. Its value is from  $\{0, 1\} $.\end{tabular}\\
\cline{2-3}
& $D$ & \begin{tabular}{@{}l@{}}Dilation. Its value is set to 1.\end{tabular}\\

\midrule
\multirow{4}*{Linear}
& $F_{\rm in}$ & \begin{tabular}{@{}l@{}} Size of input. Its value is decided by $F_{out}$ or $C_{out}$.\end{tabular}\\
\cline{2-3}
& {\color{red} $F_{\rm out}$} & \begin{tabular}{@{}l@{}}the size of output. \\Its value is decided by the number of classes \\ or $2^n$ where $n$ can be from $\{4,5,6, \cdots\}$.\end{tabular}\\

\bottomrule
\end{tabular}

}
\label{tab:hyperparameters}
\vspace{-5mm}
\end{table}

\subsection{Domain Knowledge}\label{sec:domain}

Aligned with the SOTA~\cite{maia2021can}, popular and complex network families (e.g., MLP, AlexNet, VGG and ResNet) are composed of three layer types: convolutional layer, maxpooling layer; and linear layer. Each type of layer has different layer-wise hyperparameters. We summarize them in~\autoref{tab:hyperparameters}, Note most of these hyperparameters are correlated, so that can be deterministicaly derived upon domain knowledge. The domain knowledge of each hyperparameter is summarized in~\autoref{tab:hyperparameters} and detailed in \autoref{app:domain} in Appendix.

Based on domain knowledge, \profile needs to infer three hyperparameters (i.e., $C_{\rm out}$, $K$, and $S$) for the convolutional layer, three hyperparameters (i.e., $K$,  $S$, and $P$)  for the maxpooling layer and one hyperparameter (i.e., $F_{\rm out}$) for the linear layer. When these hyperparameters are inferred by \profile, other hyperparameters can be deterministically derived from them. 
As each of these 7 hyperparameters has its own characteristics, \profile consists of 7 distinct meta-models, which are introduced in the following section.

\subsection{\profile Design}\label{sec:hymetamodel} 

For each meta-model of \profile, we customize the architecture of \segment by retaining the 4 encoders but adding a fully connected layer for either classification or regression (on whether the hyperparameter inferring is a classification or a regression task). 

As $C_{\rm out}$ and $F_{\rm out}$ have a large range of discrete values, it is almost impossible to collect the energy traces that can cover all their possible values, making it hard to infer them via classification tasks. An intuitive solution is to use regression tasks with a standard loss function, i.e., mean-square-error (MSE). However, a direct regression against either $C_{\rm out}$ or $F_{\rm out}$ is inaccurate as either one has no direct relationship with segments of sampling points predicted to their respective layer type.
We discuss how to address this challenge with a key observation later.

For other 5 hyperparameters, they have a much smaller range of value candidates and thus can be inferred via classification tasks with a standard cross-entropy (CE) loss function. During the training, 
the objective is to minimize the CE loss of ground-truth hyperparameter values and the estimated ones.
We note that when training each meta-model of \profile in the offline phase, the inputs to a model are segmented sampling points of a specific ground-truth layer type. In the online phase, the inputs to each model are a segment of sampling points predicted by \segment to a layer type. 

\mypara{Model Training for Indirect Regression Tasks}

To infer $C_{\rm out}$, we observe that for a convolutional-layer segment of sampling points, it has linear approximation with the computing overhead of a convolutional layer operations (denoted as $O_{c}$). Further, $O_{c}$ is decided as follows:
\begin{equation}\label{eq_6}
O_{c} = (C_{\rm in} \times K \times K)
\times (C_{\rm out} \times H_{\rm out} \times W_{\rm out})
\times bs.
\end{equation}
In this definition, the first bracket represents the computing overhead of a convolutional kernel that only needs to be computed once. The second bracket indicates the number of operations of the convolutional kernel. $bs$ is for batch size.
$H_{\rm out}$ and $W_{\rm out}$ respectively denote the height and width of an output feature map, which can be determined by the hyperparameters of a convolution layer as well as the height $H_{\rm in}$ and width $W_{\rm in}$ of an input feature map. We note that $H_{\rm in}$ and $W_{\rm in}$ are similar to $C_{\rm in}$ as they can be known if a preceding layer's hyperparameters are decided. 
For $H_{\rm out}$ and $W_{\rm out}$, they need to satisfy the following two equations: 
\begin{align}\label{eq_7}
& H_{\rm out} = \lfloor {(H_{\rm in} + 2 \times P - D \times (K - 1) - 1)}/{S} \rfloor + 1. \\
& W_{\rm out} = \lfloor {(W_{\rm in} + 2 \times P - D \times (K - 1) - 1)}/{S} \rfloor + 1. 
\end{align}
In these equations, $K$ and $S$ are inferred by corresponding meta-models, thus making $P$ be derived by aforementioned domain knowledge (i.e., $P = \lfloor (K - 1) / 2 \rfloor \times D$). Thus, $H_{\rm out}$ and $W_{\rm out}$ can be decided.

Further, all the parameters on the right side of 
Equation~\ref{eq_6} can be decided except $C_{\rm out}$, and we can see that $O_{c}$ is linear to $C_{\rm out}$, indicating that the regression task against $C_{\rm out}$ can be transferred to the regression against $O_{c}$. 
Specifically,  we first logarithmize $O_c$ to prevent numerical overflow, and then perform a regression task against $log(O_c)$ via minimizing the MSE between the ground-truth computing overhead and the estimated one. We then decide $C_{\rm out}$ based on Equation~\ref{eq_6}. 
Based on the aforementioned description of $C_{\rm out}$ in~\autoref{sec:domain}, $C_{\rm out}$ is inferred by rounding its regressed value 
to a nearest value in the form of $2^n$, i.e., $2^{\texttt{Round}(\log_{2}{C_{\rm out}})}$.

For $F_{\rm out}$ in the {linear} layer, we can similarly adopt the transferred solution of inferring $C_{\rm out}$ to infer $F_{\rm out}$. 
{Specifically, we define the computing overhead of a linear layer's operations (denoted as $O_{f}$) as $O_{f} = F_{\rm in} \times F_{\rm out} \times bs$. With this observation, we perform a direct regression against $O_{f}$, and then infer $F_{\rm out}$ as $F_{\rm in}$ and $bs$ are known. 
}

\section{Evaluation}\label{sec:eva}

In the online phase of \name, the target machine where a victim model is running has
Intel Xeon W 2123 (8 logical cores)
and 16\,GB DRAM. The \name attack and victim model inference run in two separate containers, managed by the underlying container scheduler. The attacker has no knowledge of the core where the victim model inference is running. 
Its system software is Ubuntu Server 18.04.6 LTS (kernel version 4.15.0-193-generic) with the latest Docker version 20.10.18 installed, where an attacker residing in one container can collect the energy traces via the \texttt{powercap} interface when 
the victim model from another container starts to perform the inference. Both containers' images are pulled from the docker hub (i.e., Ubuntu 22.04.1 LTS) and they can be different. 
The victim model uses PyTorch 1.10.1 with Python version 3.6.9 as its underlying deep learning framework. Without loss of generality, this machine is also used in the offline phase to collect the energy traces as the dataset for meta-models training. 

All meta-models are trained in a machine with an NVIDIA GeForce RTX 3070 GPU (8\,GB video memory),  Intel i7-11800H CPU (16 logical cores) and 16\,GB DRAM memory. The deep learning framework used here is PyTorch 1.10.1 with Python version 3.8.10. 

In what follows, we first describe how to construct a large-scale training/test dataset for \segment and \profile in ~\autoref{sec:datasetCon}. We then use the dataset to  train and test \segment and \profile in \autoref{sec:netrecy} and \autoref{sec:layrecy}, respectively. We note that the test dataset is used to simulate the online phase of \name so as to evaluate the attack performance of each meta-model.

\vspace{-2mm}
\subsection{Training/Testing Dataset Construction}\label{sec:datasetCon}
\vspace{-2mm}

\mypara{Sampling Energy Consumption from the RAPL} 
As the RAPL provides an energy meter via the \texttt{powercap} interface, it is accessed to sample the energy consumption for the period of a complete model inference, which is triggered by feeding a batch of images. We note that the energy consumption from two RAPL domains are collected, i.e., \texttt{PP0} and \texttt{DRAM}, corresponding to two channels of energy traces for our dataset. 
To ease the energy trace collection, \texttt{time.sleep(1)} is used to sleep for 1 second before and after the model inference. We note that \texttt{time.sleep(1)} is removed during the online phase of \name, because the attacker can trigger a victim DNN model to start inference by feeding an input and the energy trace of the model inference is distinguishable from that when the model inference is not running.

\mypara{Annotating An Energy Trace}
\segment and \profile are trained in a supervised way, thus requiring energy-trace annotation. Particularly, every sampling point in a collected energy trace is required to be annotated with a ground-truth layer-type label.
For a segment of sampling points that is labeled with a specific layer type, they are annotated with corresponding hyperparameters' values. 
As a large number of energy traces will be collected from the model inference of different model architectures, manually annotating these energy traces is extremely tedious and costly. 
To address this issue, popular deep learning frameworks do provide profiling interfaces, e.g., \texttt{torch.autograd.profiler} in PyTorch and \texttt{tf.profiler} in TensorFlow). 
In our implementation, the PyTorch-provided interface is embedded into a DNN model's deployment to automate the annotation.
We note that \name in the online  phase \textit{does not} need these interfaces.

\begin{table}[!t]
\vspace{+4mm}
\centering 
\caption{A constructed dataset for training and testing. For each model architecture, {5} different energy traces under 5 different input sizes are collected. (the number of model architectures; the number of energy traces).}
\resizebox{0.45 \textwidth}{!}
{
\begin{tabular}{c | c c c c}
\toprule
Dataset & Total & VGG  & ResNet & RandomNet \\ 
\midrule

Training
& (10,000; 50,000) & (460; 2,300)  & (4,648; 23,240) & (4,892; 24,460)\\

Test
& (1,152; 5,760) & (52; 260) & (536; 2,680) & (564; 2,820) \\

\bottomrule
\end{tabular}
}
\label{tab:model_types}
\vspace{-6mm}
\end{table}

\mypara{Constructing Training/Testing Dataset}
Aligned with the SOTA~\cite{maia2021can}, the energy traces we collect for the training/testing dataset construction are from a number of model architectures. 
Specifically, we generate model architectures from two sources. The first refers to popular and widely used model families: AlexNet, VGG, ResNet and their variants are picked here. The variant models are built  by randomly changing (i.e., adding or removing) network layers of different layer types, following the design rules of their respective model families. 
The second is to generate random models, which consist of MLPs with random combinations of only fully connected layers, and convolutional neural networks with random combinations of convolutional layers, activation layers, pooling layers, and possibly normalization layers. 

To this end, 11,152 models with different architectures have been acquired, among which
there are 512 VGG-family models, 5,184 ResNet-family models, and 5,456 randomly generated models including the structures of AlexNet family models, shown in \autoref{tab:model_types}.
We note that the model number is much larger than that of the SOTA~\cite{maia2021can}, which only considers 564 models (i.e., 500/64 for training/testing).
In our generated models, their network depth ranges from the shallowest 2 layers (i.e., 1 convolutional layer and 1 fully connected layer) to the deepest 152 layers (i.e., ResNet152). 
For the SOTA, its deepest ResNet101 has 101 layers.

Following~\cite{simonyan2014very,he2016deep,sandler2018mobilenetv2,tan2019efficientnet}, we consider 5 commonly used input sizes for each model, i.e., \{331$\times$331, 299$\times$299, 224$\times$224, 192$\times$192, 160$\times$160\}, and each one with three channels (i.e., an RGB image). The batch size is selected from [64, 96, 128, 192, 256]
\footnote{The batch size is constrained by 16\,GB memory of our experimental machine. When the input size is increasing, we decrease the maximum batch size to not exceed the memory capacity.}. 
In total, 55,760 energy traces are collected for the 11,152 different models (each of the five input sizes is applied to a same model), resulting in {20.4\,GB}. 
Considering the proportion of the models in each of the three model families, 5,760 energy traces are randomly chosen as the test dataset, and the rest are used as the training dataset.

\vspace{-2mm}
\subsection{Network Structure Recovery}\label{sec:netrecy}
\vspace{-2mm}

To evaluate the structure recovering performance of \segment, we use two metrics, i.e., Levenshtein Distance Accuracy (LDA) and Segment Accuracy (SA), following the SOTA work~\cite{maia2021can}.

\vspace{2pt}\noindent$\bullet$ {LDA} is the similarity between a predicted structure and a ground-truth structure\footnote{Generally, LDA is a metric related to Normalised Levenstein Distance (NLD)~\cite{graves2006connectionist,hu2020deepsniffer,maia2021can}. 
The Levenstein Distance is a measure of the minimum number of edits required to transform one sequence into another. The allowed edits include insertions, deletions, and substitutions. Thus, LDA is equal to $\{1 - {\rm NLD}\}$ where NLD shows the error rate.}. 

\vspace{2pt}\noindent$\bullet$ {SA} is the percentage of sampling points that are correctly predicted into their ground-truth layer types.

For a model stealing attack, LDA indicates its prediction accuracy of a network structure. It is 100\% when a predicated structure exactly matches a ground-truth one. SA measures the segmentation performance of \segment. Clearly, a higher SA, better performance of \profile.

We use the SGD optimizer to train \segment at a learning rate of $0.01$ in conjunction with a cosine annealing learning rate. The epochs are 100 with a batch size of 32. Before each batch is fed into the model, energy traces in the batch are padded to the length of the longest energy trace in the batch so that the energy traces within the batch are of equal length. This padding operation is equivalent to a potential data enhancement, similar to a random crop in an image.

\begin{figure}[!t]
	\centering
	\includegraphics[trim=0 0 0 0,clip,width=0.35 \textwidth]{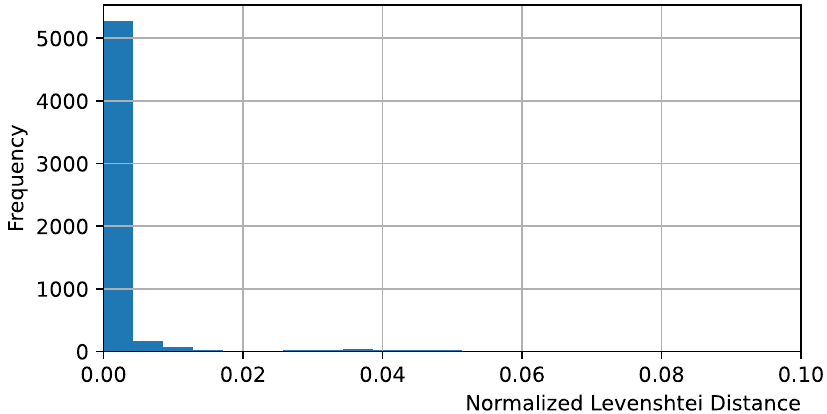}
	\caption{A distribution of normalized Levenshtein distances on the testing dataset of 5,760  energy traces. 0 means an exact network-structure match.}
	\label{fig:hist}
\end{figure}

The performance in terms of LDA and SA of \segment on the testing dataset is shown in~\autoref{tab:result}.
Specifically, LDA for all the layer types are more than 98\%. 
For SA, \name performs better in recovering layers that require intensive computation (i.e., convolutional layer) than those layers with less intensive computation (e.g., Average Pooling and Linear). This is because the layers with less intensive computation correspond to a smaller number of sampling points.
Despite this, SA of all layer types is still greater than 92\%.
Overall, \segment has achieved 99.26\% for SA and 99.75\% for LDA, respectively.  
Additionally, we plot the distribution of normalized Levenshtein distances for all the testing energy traces in \autoref{fig:hist}.  We see that almost all of them have 0 normalized Levenshtein distances, indicating that the estimated network structures have exactly matched the ground-truth ones. 

\begin{table}[!t]
\vspace{+4mm}
\centering 
\caption{The performance of \segment in recovering network structures in the first-step of \name.}
\resizebox{0.5 \textwidth}{!}
{
\begin{threeparttable}
\begin{tabular}{c | c c c c c c c c}
\toprule
{\bf Layer} & Conv & BN & ReLu & {\begin{tabular}[c]{@{}c@{}}MP \\ (Max. Pool.) \end{tabular}} & {\begin{tabular}[c]{@{}c@{}} AP \\ (Ave. Pool.)\end{tabular}} & Linear & Add$^*$ & Overall \\
\midrule

SA (\%) & 99.44 & 97.99 & 98.02 & 99.57 & 92.93 & 94.43 & 95.70 & 99.26\\
\midrule

LDA (\%) & 99.85 & 99.74 & 99.76 & 99.69 & 99.06 & 98.47 & 99.64 & 99.75 \\

\bottomrule
\end{tabular}
\begin{tablenotes}
      \item $^*$Add is a vector addition operation in residual shortcut concatenation.
\end{tablenotes}
\end{threeparttable}

}
\label{tab:result}
\vspace{-6mm}
\end{table}

\vspace{-2mm}
\subsection{Layer-Wise Hyperparameter Inferring}\label{sec:layrecy}
\vspace{-2mm}

A segment of sampling points for a predicated layer is fed to a corresponding meta-model of \profile to infer the layer-wise hyperparameter value. 
The configuration (e.g., SGD optimizer, learning rate) of training \profile is the same as \segment. Before each segment of sampling points as an input is fed into a meta-model, it is normalized and then uniformly resized to a specified length of 1024 sampling points, which is the average length of all the segments of sampling points.

We then try to evaluate the performance of each meta-model in \profile using three metrics, i.e., prediction, recall, and F1 score. \autoref{tab:hresult} shows these metrics for the inferred hyperparameters of different layer types. Clearly, all the hyperparameters in the table are accurately inferred, i.e., all the scores in each metric are almost 100\%, which is largely attributed to the  precise segmentation of \segment.
For the challenges of inferring $C_{\rm out}$ and $F_{out}$, their scores reach about 99\%, which have validated our proposed indirect regression (see \autoref{sec:hymetamodel}). 
Further, we compare the inferring performance of $C_{\rm out}$ using our indirect regression with that of an intuitive direct regression in \autoref{sec:dis}, results of which show our approach performs much better than the direct one.

\vspace{-2mm}
\subsection{Comparing \name with the SOTA}\label{sec:compressionComp}
\vspace{-2mm}

Despite the SOTA attack~\cite{maia2021can} is inapplicable to the MLaaS setting,
we compare \name with it for two main reasons. 
First, both works are based on a learning-based framework to steal model architectures.
Second, different from other existing works~\cite{wei2020leaky,yan2020cache}, both have evaluated a large number of large and deep networks, which are prevalent in the real-world. 

\mypara{Qualitative Comparison} 
While the SOTA~\cite{maia2021can} has open-sourced its code\footnote{The source code of~\cite{maia2021can} is available at \url{https://github.com/henriquetmaia/gpuNetSnooper}.}, its training/test dataset of EM/magnetic signals is not publicly available.
Hence, we make a qualitative comparison based on the reported results in~\cite{maia2021can}. While \name has a much lower sampling rate than~\cite{maia2021can} (i.e., 1\,KHz vs 47\,KHz),
it has overall 99.26\% SA and 99.75\% LDA for the network structure (i.e., network topology in ~\cite{maia2021can}), which are clearly better than the SOTA's overall 96.8\% SA and 88.2\% LDA. 

For the performance of inferring convolutional layer-wise hyperparameters, the SOTA has tested 1,804 convolutional layers and the accuracy of the layer-wise hyperparameters is between 96\%-97\% in Table S2 of~\cite{maia2021can}. Additionally, the SOTA has not shown the accuracy of inferring the kernel number for a convolutional layer, a challenge that has been addressed by \name. 
Compared to it, \name has tested 226,235 convolutional layers and achieved an accuracy of more than 98\% (shown in \autoref{tab:hresult}), including the performance of inferring the kernel number. 
We note that the SOTA did not provide the statistics of inferring hyperparameters of other layer types (i.e., max pooling layer and linear layer), which thus we cannot compare with \name.

\begin{table}[!t]
\vspace{+4mm}
\centering 
\caption{Inferring performance of each hyperparameter in the second-step of \name. 
}
\resizebox{0.40\textwidth}{!}
{
\begin{tabular}{c | c c c c}
\toprule
Layer & Hyperparameter & Precision(\%) & Recall(\%) & F1(\%)\\ 

\midrule
\multirow{3}*{Convolutional}
& $C_{\rm out}$ & 98.89 & 98.98 & 98.93\\
& $K$ & 99.86 & 99.98 & 99.92\\
& $S$ & 99.87 & 99.77 & 99.82\\

\midrule
\multirow{3}*{MaxPooling}
& $K$ & 100 & 100 & 100 \\
& $S$ & 100 & 100 & 100 \\
& $P$ & 100 & 100 & 100 \\

\midrule
\multirow{1}*{Linear}
& $F_{out}$ & 99.18 & 99.04 & 99.10 \\

\midrule
\midrule
\multicolumn{2}{c}{Weighted average}
& 99.58 & 99.61 & 99.60 \\

\bottomrule
\end{tabular}
}
\label{tab:hresult}
\end{table}

\begin{figure}
	\centering
	\includegraphics[trim=0 0 0 0,clip,width=0.45 \textwidth]{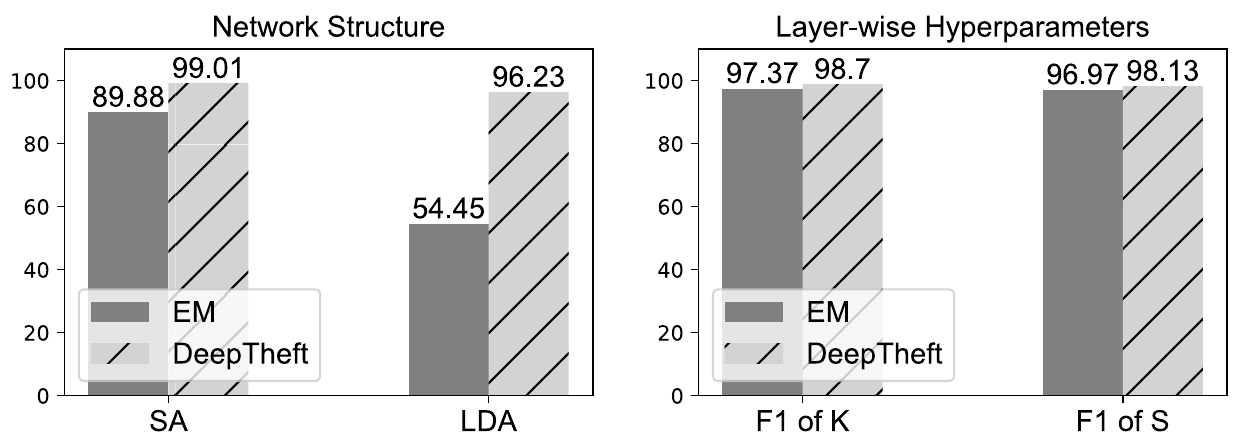}
	\caption{A comparison between \name and $EM$
 (The $EM$ refers to the SOTA attack~\cite{maia2021can}).}
	\label{fig:bar}
    \vspace{-6mm}
\end{figure}

\mypara{Quantitative Comparison} 
Considering that the first-step is critical to the second-step in both works, we further make a quantitative comparison between them.
Specifically, for the first-step. we rely on the SOTA's open-source code to reproduce its proposed BiLSTM model for segmenting our energy traces. For the second-step, both works use the set of meta-models, i.e., \profile. 
For the training dataset, the SOTA~\cite{maia2021can} uses 500 models only. Thus, we use the same model number by randomly picking 500 from the 10,000 models in our training dataset. 
To evaluate the attack performance of the SOTA and \name, our testing dataset consisting of 1,152 architectures is applied. 

The evaluation results are shown in \autoref{fig:bar}. For network structure recovery, the SOTA has only 54.45\% LDA, and \name achieves a much higher LDA of 96.23\%, indicating that \segment has retained its performance even when the size of the training dataset has decreased significantly from 10,000 to 500.
For layer-wise hyperparameters inferring,  \name is slightly better than EM.
We note that for both works, {only those sampling points that are predicted into correct layer types in the first-step are fed into the second step} where our proposed \profile is in use. Thus, while the SOTA shows 89.88\% SA, lower than \name that has 99.01\% SA in the first step, the SOTA assisted by \profile in the second step performs much better, i.e., showing higher performance and getting closer to \name. This demonstrates the efficacy of \profile in inferring layer-wise hyperparameters, and also implies that \profile can tolerate a relatively low SA to an extent.

\mypara{Training Overhead} 
As BiLSTM in the SOTA processes the sampling points in a serial way, its execution latency grows linearly with the length of time-series sampling points. Thus, it is time-consuming for BiLSTM to perform training and inference against a long sequence of sampling points. 
In comparison, \name leverages CNN, which processes all the sampling points per energy trace in a concurrent way, resulting in significantly less execution latency.
We evaluate the training overhead of BiLSTM and \segment using the aforementioned small training dataset of 500 random models, results of which show that \segment takes only half a minute per epoch, much less than BiLSTM which requires about 13 minutes per epoch.

\vspace{-2mm}
\subsection{\name Generalization}\label{sec:generic}
\vspace{-2mm}

\name can be generic to steal model architecture through other side channels that generate time-series signals. Specifically, we validate that a CPU-frequency side channel~\cite{wang2022hertzbleed}, also for the first time, can be utilized to accurately steal a target model architecture via our proposed learning framework. 
For the side channel, it exists in modern x86 processors, where either local or remote attackers could observe DVFS-induced frequency variations depend on the current data being processed, and thus break the data confidentiality of cryptographic software.

In the aforementioned test machine, the same Ubuntu server is running as a bare-metal OS. Within this server, we initiate two unprivileged processes running \name and a model inference, respectively.
Similar to the dataset construction in~\autoref{sec:datasetCon}, \name leverages an open-source toolkit\footnote{ \url{https://github.com/FPSG-UIUC/hertzbleed}} to collect CPU-frequency traces at a sampling rate of 1\,KHz, which costs two weeks and generates a data volume of 6.8\,GB. The data volume consists of 55,760 frequency traces, 
among which 50,000 and 5760 are used as the training and testing dataset, respectively.
Similarly, the traces are from forward inference of the 11,152 different model architectures with the five different input sizes. 

While the CPU frequency trace looks significantly different from that of the power side channel (a CPU-frequency trace example is shown in \autoref{fig:power_freq} of Appendix), 
our experimental results of \name with CPU-frequency traces shows LDA and SA of 99.04\% and 99.20\%, which are similar to that of using \name with power traces.

\section{Discussion}\label{sec:dis}
We first evaluate the impacts of input size and RAPL domains on \name. We then perform extensive ablation studies on the learning framework. Moreover, we evaluate the impact of noisy workloads on the \name performance. We further discuss the threat of \name to commercial clouds. Last, some countermeasures are discussed.

\vspace{2mm}
\mypara{Input Size Impact on \name}
In the above experiments, 5 commonly input sizes are used for models in the training/test dataset. To evaluate the impact of a specified input size on the attack performance of \name, we also show the accuracy of recovering network structure and layer-wise hyperparameters for each of the 5 aforementioned input sizes.
As shown in \autoref{tab:sizeresult}, both overall LDA (indicating the accuracy of network structure recovery) and F1 (showing the accuracy of inferring layer-wise hyperparameters) are similar to each other in different input sizes, indicating that \name is insensitive to the input size.

\begin{table}[!h]
\vspace{-2mm}
\centering 
\caption{\name's  performance is insensitive to the input sizes (batch size, height, width, channel number).}
\resizebox{0.35 \textwidth}{!}
{
\begin{tabular}{c | c c}
\toprule
Input Size & Overall LDA(\%) & Ave. F1(\%)\\ 

\midrule
(256, 160, 160, 3) & 99.65 & 99.63 \\
(192, 192, 192, 3) & 99.72 & 99.59 \\
(128, 224, 224, 3) & 99.81 & 99.58 \\
(96, 299, 299, 3) & 99.73 & 99.62 \\
(64, 331, 331, 3) & 99.85 & 99.61 \\

\bottomrule
\end{tabular}
}
\label{tab:sizeresult}
\vspace{-2mm}
\end{table}

\mypara{RAPL Domains Impact on \name}
In the above experiments, both \texttt{PP0} and \texttt{DRAM} domains provided by the RAPL are used. Here, we compare the attack performance of \name using both domains with that of using an individual domain.
As shown in \autoref{tab:channelresult}, using a single domain can achieve a high recovering accuracy, which can be attributed to our proposed meta-models.
Compared to each domain used, using both has slightly better accuracy.

\begin{table}[h]
\vspace{-2mm}
\centering 
\caption{\name's performance is slightly better when both RAPL domains are used. 
}
\resizebox{0.3 \textwidth}{!}
{
\begin{tabular}{c | c c}
\toprule
RAPL Domain & Overall LDA(\%) & Ave. F1(\%)\\ 

\midrule
PP0 & 98.07 & 98.72 \\
DRAM & 97.79 &  98.88 \\
PP0 + DRAM & 99.75 & 99.60 \\

\bottomrule
\end{tabular}
}
\label{tab:channelresult}
\vspace{-4mm}
\end{table}

\vspace{2mm}
\mypara{Direct Regression Against Kernel Number}
As discussed in \autoref{sec:hymetamodel}, performing a direct regression task against the kernel number ($C_{\rm out}$) is not accurate and instead we have proposed an indirect regression approach. 
Here, \autoref{tab:coutresult} shows the performance of inferring $C_{\rm out}$ between direct regression and our proposed indirect regression. 
Clearly, the indirect regression is better in the three evaluated metrics than the direct regression in the default training-dataset size of 50,000.
We also reduce the size of the training dataset by 20 times to show its impact on the two methods. 
For the direct regression, it experiences an obvious performance drop in the decreased size of 2,500 energy traces, which is unacceptable for inferring $C_{\rm out}$.
For the indirect regression, its performance is almost not affected by the decreased size, showing its robustness in inferring $C_{\rm out}$.

\begin{table}[!t]
\vspace{4mm}
\centering 
\caption{Inferring performance of the kernel number using direct or indirect regression of \profile.
}
\resizebox{0.45 \textwidth}{!}
{
\begin{tabular}{c | c c c c}
\toprule
Training Size & Inferring Method & Precision(\%) & Recall(\%) & F1(\%)\\ 

\midrule
\multirow{2}*{50,000 traces} 
& Direct Regression & 95.57 & 91.55 & 93.28 \\
& Indirect Regression & 98.89 & 98.98 & 98.93 \\

\midrule
\multirow{2}*{2,500 traces} 
& Direct Regression & 77.58 & 70.12 & 72.54 \\
& Indirect Regression & 97.02 & 97.20 & 97.09 \\

\bottomrule
\end{tabular}
}
\label{tab:coutresult}
 \vspace{-5mm}
\end{table}

\vspace{2mm}
\mypara{Ablation Study}
We perform an ablation study of our proposed learning framework. Particularly, to show its attacking effectiveness, we remove its critical components and losses.
As mentioned in Section~\ref{sec:compressionComp}, \name achieves SA of 99.01\% and LDA of 96.23\%. If only BiLSTM is kept, SA and LDA have dropped to 89.88\% and 54.45\%, respectively. If only UNet is retained, SA and LDA have decreased to 96.57\% and 58.16\%.
If only Temporal-Encoder is removed, SA and LDA becomes 98.32\% and 78.19\%. If only UPloss is removed, 
SA and LDA are 98.03\% and 83.58\%. 

\vspace{2mm}
\mypara{Cross-CPU} Previous experiments are aligned with all existing works’ assumptions where targeted machine configurations (e.g., CPU) are known to an attacker. So that all training energy traces are collected from the same CPU as the targeted machine. We have further collected energy traces from another Intel i7-4710MQ and immediately evaluated it (without transfer learning) with meta-models trained from Intel Xeon-W-2123 (previous experiments are upon Intel Xeon-W-2123). As expected, the SA/LDA becomes 79.79\%/59.83\% due to energy trace variations across CPUs. To improve the attack performance, we then leverage a few new traces (100 model architectures’ energy traces with five input size per model from Intel i7-4710MQ) to perform transfer learning upon meta-models obtained through Intel Xeon-W-2123. With transfer learning, the SA/LDA has been improved to 99.07\%/97.55\%.

\vspace{2mm}
\mypara{Noisy Workload}
To evaluate the impact of noisy workloads, a third container is created to host a running LAMP (Linux, Apache, MySQL and PHP/Perl/Python) server with its default configuration. It co-resides with two other containers hosting \name and a victim model respectively. 
In this setting, 500 energy traces of 100 random architectures (five input-sizes per architecture) are collected as a testing dataset against the training dataset generated in Section~\ref{sec:eva}.
As a baseline, in the setting where there is no such noise, another 500 energy traces against the same random architectures are collected.
Because of the noise, the pair of SA and LDA has insignificantly dropped to 98.95\% and 94.77\% compared to the baseline pair of 99.07\% and 98.13\%. 

The accuracy drop is expected to be compensated given repeated queries on the same victim model architecture---note previous \name performance is upon a single trace snapshot. To this end, we have repeated three times (could be more) on each of 100 different architectures per input size (five input sizes).
Algorithms of i) pre-averaging repeated traces and ii) post-majority-voting results are considered.
The former improves SA and LDA to 99.03\% and 96.68\%, respectively. The latter (slightly) outperforms the former, presenting a 99.09\% SA and 96.91\%.

\vspace{2mm}
\mypara{Threat to Commercial Clouds}
Since the RAPL-based power side-channel has been discovered by Lipp et al.~\cite{lipp2021platypus}, unprivileged access to the \texttt{powercap} interface in bare-metal OS has been disabled. 
Despite that, some container-based clouds expose the \texttt{powercap} to a container~\cite{gao2017containerleaks,zhang2021red}. Based on our experiments, a container, in the latest docker environment of 20.10.18, can access the \texttt{powercap} without limitation. 
Additionally, there exist leading cloud providers that \emph{do} allow a VM to access the RAPL MSRs, e.g., AWS EC2 reported by Zhang et al.~\cite{zhang2021red}.

\begin{figure}[h!]
	\centering
	\includegraphics[trim=0 0 0 0,clip,width=0.40\textwidth]{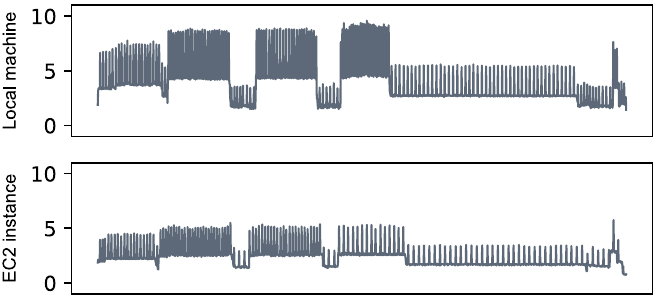}
	\caption{Two energy consumption traces from the \texttt{DRAM} domain are collected for the same model architecture via our experimental local machine (top) and one EC2 instance of c4.8xlarge (bottom).\vspace{-0.7cm}}
	\label{fig:power_comparison}
	\vspace{+4mm}
\end{figure}

Following their observation, we have further confirmed that some (not all) EC2 instances have provided access to RAPL.
Specifically, some EC2 instances are equipped with a tool called \texttt{turbostat} to help monitor the operating state of a processor, such as its frequency, C-state (referring to controlling the sleep levels that a core can enter when it is idle), and P-state (referring to controlling the desired performance in CPU frequency from a core). 

More importantly, an EC2 instance can leverage this tool to access RAPL MSRs and thus acquire energy consumption from two domains, i.e., \texttt{Package} domain and \texttt{DRAM} domain. 
As AWS provides users with various EC2 instance types, not all of them have provided \texttt{turbostat} to access RAPL MSRs.
We have experimented with a number of EC2 instances and~\autoref{tab:ec2} shows whether each instance type supports access to RAPL MSRs.
A list of AWS EC2 instances that might enable/disable the RAPL access can be found here\footnote{\url{https://docs.aws.amazon.com/AWSEC2/latest/UserGuide/processor_state_control.html}}.

\begin{table}[h]
\vspace{4mm}
\centering 
\caption{Based on our experiments, some AWS EC2 instance types are listed to show whether they are enabled/disabled to access RAPL MSRs.}
\resizebox{0.45 \textwidth}{!}
{
\begin{tabular}{c | c c}
\toprule
AWS EC2 & Enabled & Disabled\\ 

\midrule
\begin{tabular}{c}Instance Types \end{tabular} & \begin{tabular}{c}c4.8xlarge, r4.8xlarge, \\d2.8xlarge, i3.8xlarge, \\x1e.8xlarge, h1.8xlarge\end{tabular} 
 & \begin{tabular}{c}t1.micro, t2.xlarge, \\t3.xlarge, c3.2xlarge, \\c4.2xlarge, c5.2xlarge\end{tabular} \\

\bottomrule
\end{tabular}
}

\label{tab:ec2}
\end{table}

On top of that, we pick one instance type of c4.8xlarge to collect the energy trace {from the \texttt{DRAM} domain} for a given model and compare it with that of our experimental local machine in~\autoref{sec:eva}, results of which are shown in~\autoref{fig:power_comparison}. As can be seen from this figure, their energy traces are similar to each other in terms of their trace shapes,  
indicating that \name, running within an EC2 instance supports the RAPL access, can steal a targeted model architecture running within another EC2 instance. 
To be specific, for an instance running model inference from an unknown model, \name can collect its energy consumption traces from another instance. We note that both instances are assumed to co-reside in the same physical machine and only the instance running \name needs access to RAPL MSRs.

\vspace{2mm}
\mypara{Mitigation}
First, limiting container-level access to the RAPL interface can mitigate the power side-channel utilized by Deeptheft. Second, disabling either Turbo Boost or SpeedStep/Hardware Controlled Performance States from the BIOS can mitigate the frequency side-channel exploited by DeepTheft. However, given that our learning-framework is generic to time-series side-channel signals (low sampling rate in particular) besides the validated power and frequency side-channels, more general side-channel mitigation techniques such as masking hardware protection and randomization (e.g., noise injection) are demanded and can be incorporated during the model inference.

As for masking, it divides sensitive information into multiple shares. One of the primary assumptions in masking is that the leakage of each share is independent of the other shares. A successful attack by utilizing the leakage from a single share cannot be performed unless the leakage information from all of the shares is available. Effectively incorporating masking into model inference is an interesting future work. Injecting delicate heavy background noise (e.g., one common means of randomization) is a common approach to reducing the signal-to-noise ratio (SNR) to mitigate side-channel attacks. We have evaluated this by significantly increasing CPU workload: a stress command (\texttt{stress -m 8 --vm-bytes 1G}) has been utilized to induce high system noise. Consequently, the SA/LDA decreases from 99.26\%/99.75\% to 71.28\%/46.71\%. However, we note that repeated measurements can enhance the SNR, effectively reducing the injected noise.

\section{Conclusion}\label{sec:conclusion}
\name is a new end-to-end model architecture stealing attack that leverages time-series side-channel signals (e.g., power and CPU frequency are used in our paper) in the MLaaS setting. 
Given the constraint of a low sampling rate, we proposed a new and generic supervised-learning framework that consists of a set of meta-models, i.e., \segment and \profile, based on which \name can recover a complete network structure and layer-wise hyperparameters.
Our extensive experiments have validated the \name's high efficacy in both model structure topology recovery and layer-wise hyperparameters inferring under a low sampling rate.

\bibliographystyle{acm}
\bibliography{defs,refs}

\begin{thebibliography}{10}

\bibitem{AlibabaCPU}
{\sc {Alibaba Cloud}}.
\newblock {ECS Bare Metal Instance Types}.
\newblock
  \url{https://www.alibabacloud.com/help/en/elastic-compute-service/latest/ecs-bare-metal-instance-types-overview},
  2023.

\bibitem{awsmlaas}
{\sc {Amazon, Inc.}}
\newblock Amazon machine learning.
\newblock \url{https://aws. amazon.com/machine-learning}, 2018.

\bibitem{AWSCPU}
{\sc {Amazon, Inc.}}
\newblock {AWS CPU Inference}.
\newblock
  \url{https://docs.aws.amazon.com/deep-learning-containers/latest/devguide/deep-learning-containers-eks-tutorials-cpu-inference.html},
  2018.

\bibitem{AWSdocker}
{\sc {Amazon, Inc.}}
\newblock {AWS Deep Learning Containers}.
\newblock
  \url{https://docs.aws.amazon.com/deep-learning-containers/latest/devguide/deep-learning-containers-ec2-tutorials-inference.html},
  accessed: 11-Oct-2022.

\bibitem{atya2017malicious}
{\sc Atya, A. O.~F., Qian, Z., Krishnamurthy, S.~V., La~Porta, T., McDaniel,
  P., and Marvel, L.}
\newblock Malicious co-residency on the cloud: Attacks and defense.
\newblock In {\em INFOCOM\/} (2017), pp.~1--9.

\bibitem{balle2022reconstructing}
{\sc Balle, B., Cherubin, G., and Hayes, J.}
\newblock Reconstructing training data with informed adversaries.
\newblock In {\em S\&P\/} (2022), IEEE Computer Society, pp.~1556--1556.

\bibitem{batina2019csi}
{\sc Batina, L., Bhasin, S., Jap, D., and Picek, S.}
\newblock {CSI NN}: Reverse engineering of neural network architectures through
  electromagnetic side channel.
\newblock In {\em Usenix Security\/} (2019), pp.~515--532.

\bibitem{carlini2022membership}
{\sc Carlini, N., Chien, S., Nasr, M., Song, S., Terzis, A., and Tramer, F.}
\newblock Membership inference attacks from first principles.
\newblock In {\em S\&P\/} (2022), IEEE, pp.~1897--1914.

\bibitem{carlini2020cryptanalytic}
{\sc Carlini, N., Jagielski, M., and Mironov, I.}
\newblock Cryptanalytic extraction of neural network models.
\newblock In {\em CRYPTO\/} (2020).

\bibitem{chmielewski2021reverse}
{\sc Chmielewski, {\L}., and Weissbart, L.}
\newblock On reverse engineering neural network implementation on {GPU}.
\newblock In {\em ACNS\/} (2021).

\bibitem{docker}
{\sc {Docker, Inc.}}
\newblock {Home - Docker}.
\newblock \url{https://www.docker.com/}, 2013.

\bibitem{ezhilchelvan2015evaluating}
{\sc Ezhilchelvan, P.~D., and Mitrani, I.}
\newblock Evaluating the probability of malicious co-residency in public
  clouds.
\newblock {\em IEEE Transactions on Cloud Computing\/} (2015).

\bibitem{fredrikson2015model}
{\sc Fredrikson, M., Jha, S., and Ristenpart, T.}
\newblock Model inversion attacks that exploit confidence information and basic
  countermeasures.
\newblock In {\em Proc. CCS\/} (2015).

\bibitem{ganju2018property}
{\sc Ganju, K., Wang, Q., Yang, W., Gunter, C.~A., and Borisov, N.}
\newblock Property inference attacks on fully connected neural networks using
  permutation invariant representations.
\newblock In {\em Proc. CCS\/} (2018), pp.~619--633.

\bibitem{gao2017containerleaks}
{\sc Gao, X., Gu, Z., Kayaalp, M., Pendarakis, D., and Wang, H.}
\newblock Containerleaks: Emerging security threats of information leakages in
  container clouds.
\newblock In {\em DSN\/} (2017).

\bibitem{GoogleCPU}
{\sc {Google}}.
\newblock {CPU platforms}.
\newblock \url{https://cloud.google.com/compute/docs/cpu-platforms}, 2023.

\bibitem{googlemlaas}
{\sc {Google, Inc.}}
\newblock Cloud ml engine overview.
\newblock \url{https://cloud.google.com/ml-engine/docs/technical-overview},
  2018.

\bibitem{graves2006connectionist}
{\sc Graves, A., Fern{\'a}ndez, S., Gomez, F., and Schmidhuber, J.}
\newblock Connectionist temporal classification: labelling unsegmented sequence
  data with recurrent neural networks.
\newblock In {\em Proc. ICML\/} (2006), pp.~369--376.

\bibitem{graves2005bidirectional}
{\sc Graves, A., Fern{\'a}ndez, S., and Schmidhuber, J.}
\newblock Bidirectional {LSTM} networks for improved phoneme classification and
  recognition.
\newblock In {\em ICANN\/} (2005), pp.~799--804.

\bibitem{guliani2019per}
{\sc Guliani, A., and Swift, M.~M.}
\newblock Per-application power delivery.
\newblock In {\em EUROSYS\/} (2019), pp.~1--16.

\bibitem{hazelwood2018applied}
{\sc Hazelwood, K., Bird, S., Brooks, D., Chintala, S., Diril, U., Dzhulgakov,
  D., Fawzy, M., Jia, B., Jia, Y., Kalro, A., et~al.}
\newblock Applied machine learning at facebook: A datacenter infrastructure
  perspective.
\newblock In {\em HPCA\/} (2018), pp.~620--629.

\bibitem{he2016deep}
{\sc He, K., Zhang, X., Ren, S., and Sun, J.}
\newblock Deep residual learning for image recognition.
\newblock In {\em Proc. CVPR\/} (2016), pp.~770--778.

\bibitem{hu2020deepsniffer}
{\sc Hu, X., Liang, L., Li, S., Deng, L., Zuo, P., Ji, Y., Xie, X., Ding, Y.,
  Liu, C., Sherwood, T., et~al.}
\newblock Deepsniffer: A {DNN} model extraction framework based on learning
  architectural hints.
\newblock In {\em ASPLOS\/} (2020), pp.~385--399.

\bibitem{hua2018reverse}
{\sc Hua, W., Zhang, Z., and Suh, G.~E.}
\newblock Reverse engineering convolutional neural networks through
  side-channel information leaks.
\newblock In {\em DAC\/} (2018), pp.~1--6.

\bibitem{liang2022clairvoyance}
{\sc Liang, S., Zhan, Z., Yao, F., Cheng, L., and Zhang, Z.}
\newblock {Clairvoyance}: Exploiting far-field {EM} emanations of {GPU} to
  ``see" your {DNN} models through obstacles at a distance.
\newblock In {\em IEEE S\&P Workshops\/} (2022), pp.~312--322.

\bibitem{lipp2022amd}
{\sc Lipp, M., Gruss, D., and Schwarz, M.}
\newblock Amd prefetch attacks through power and time.
\newblock In {\em USENIX Security\/} (2022).

\bibitem{lipp2021platypus}
{\sc Lipp, M., Kogler, A., Oswald, D., Schwarz, M., Easdon, C., Canella, C.,
  and Gruss, D.}
\newblock Platypus: Software-based power side-channel attacks on x86.
\newblock In {\em S\&P\/} (2021), pp.~355--371.

\bibitem{maia2021can}
{\sc Maia, H.~T., Xiao, C., Li, D., Grinspun, E., and Zheng, C.}
\newblock Can one hear the shape of a neural network? snooping the {GPU} via
  magnetic side channel.
\newblock In {\em Usenix Security\/} (2022).

\bibitem{meyers2022reverse}
{\sc Meyers, V., Gnad, D., and Tahoori, M.}
\newblock Reverse engineering neural network folding with remote {FPGA} power
  analysis.
\newblock In {\em IEEE Annual International Symposium on Field-Programmable
  Custom Computing Machines\/} (2022), pp.~1--10.

\bibitem{naghibijouybari2018rendered}
{\sc Naghibijouybari, H., Neupane, A., Qian, Z., and Abu-Ghazaleh, N.}
\newblock Rendered insecure: Gpu side channel attacks are practical.
\newblock In {\em CCS\/} (2018), pp.~2139--2153.

\bibitem{oliynyk2022know}
{\sc Oliynyk, D., Mayer, R., and Rauber, A.}
\newblock I know what you trained last summer: A survey on stealing machine
  learning models and defences.
\newblock {\em arXiv preprint arXiv:2206.08451\/} (2022).

\bibitem{rakin2022deepsteal}
{\sc Rakin, A.~S., Chowdhuryy, M. H.~I., Yao, F., and Fan, D.}
\newblock {DeepSteal}: Advanced model extractions leveraging efficient weight
  stealing in memories.
\newblock In {\em S\&P\/} (2022), pp.~1157--1174.

\bibitem{ribeiro2015mlaas}
{\sc Ribeiro, M., Grolinger, K., and Capretz, M.~A.}
\newblock Mlaas: Machine learning as a service.
\newblock In {\em ICMLA\/} (2015), pp.~896--902.

\bibitem{ristenpart2009hey}
{\sc Ristenpart, T., Tromer, E., Shacham, H., and Savage, S.}
\newblock Hey, you, get off of my cloud: exploring information leakage in
  third-party compute clouds.
\newblock In {\em CCS\/} (2009), pp.~199--212.

\bibitem{ronneberger2015u}
{\sc Ronneberger, O., Fischer, P., and Brox, T.}
\newblock U-net: Convolutional networks for biomedical image segmentation.
\newblock In {\em MICCAI\/} (2015), pp.~234--241.

\bibitem{sandler2018mobilenetv2}
{\sc Sandler, M., Howard, A., Zhu, M., Zhmoginov, A., and Chen, L.-C.}
\newblock Mobilenetv2: Inverted residuals and linear bottlenecks.
\newblock In {\em Proc. CVPR\/} (2018), pp.~4510--4520.

\bibitem{sarood2013optimizing}
{\sc Sarood, O., Langer, A., Kal{\'e}, L., Rountree, B., and De~Supinski, B.}
\newblock Optimizing power allocation to {CPU} and memory subsystems in
  overprovisioned {HPC} systems.
\newblock In {\em IEEE International Conference on Cluster Computing\/} (2013),
  pp.~1--8.

\bibitem{shokri2017membership}
{\sc Shokri, R., Stronati, M., Song, C., and Shmatikov, V.}
\newblock Membership inference attacks against machine learning models.
\newblock In {\em S\&P\/} (2017), pp.~3--18.

\bibitem{simonyan2014very}
{\sc Simonyan, K., and Zisserman, A.}
\newblock Very deep convolutional networks for large-scale image recognition.
\newblock {\em arXiv preprint arXiv:1409.1556\/} (2014).

\bibitem{tan2019efficientnet}
{\sc Tan, M., and Le, Q.}
\newblock Efficientnet: Rethinking model scaling for convolutional neural
  networks.
\newblock In {\em ICML\/} (2019), PMLR, pp.~6105--6114.

\bibitem{tian2021remote}
{\sc Tian, S., Moini, S., Wolnikowski, A., Holcomb, D., Tessier, R., and
  Szefer, J.}
\newblock Remote power attacks on the versatile tensor accelerator in
  multi-tenant {FPGAs}.
\newblock In {\em FCCM\/} (2021), pp.~242--246.

\bibitem{tramer2016stealing}
{\sc Tram{\`e}r, F., Zhang, F., Juels, A., Reiter, M.~K., and Ristenpart, T.}
\newblock Stealing machine learning models via prediction {APIs}.
\newblock In {\em Usenix Security\/} (2016), pp.~601--618.

\bibitem{varadarajan2015placement}
{\sc Varadarajan, V., Zhang, Y., Ristenpart, T., and Swift, M.}
\newblock A placement vulnerability study in $\{$Multi-Tenant$\}$ public
  clouds.
\newblock In {\em Usenix Security\/} (2015), pp.~913--928.

\bibitem{wang2018stealing}
{\sc Wang, B., and Gong, N.~Z.}
\newblock Stealing hyperparameters in machine learning.
\newblock In {\em S\&P\/} (2018), IEEE, pp.~36--52.

\bibitem{wang2022hertzbleed}
{\sc Wang, Y., Paccagnella, R., He, E.~T., Shacham, H., Fletcher, C.~W., and
  Kohlbrenner, D.}
\newblock Hertzbleed: Turning power side-channel attacks into remote timing
  attacks on x86.
\newblock In {\em USENIX Security\/} (2022), pp.~679--697.

\bibitem{wei2020leaky}
{\sc Wei, J., Zhang, Y., Zhou, Z., Li, Z., and Al~Faruque, M.~A.}
\newblock Leaky {DNN}: Stealing deep-learning model secret with {GPU}
  context-switching side-channel.
\newblock In {\em DSN\/} (2020), pp.~125--137.

\bibitem{wei2018know}
{\sc Wei, L., Luo, B., Li, Y., Liu, Y., and Xu, Q.}
\newblock I know what you see: Power side-channel attack on convolutional
  neural network accelerators.
\newblock In {\em ACSAC\/} (2018), pp.~393--406.

\bibitem{wolf2021stealing}
{\sc Wolf, S., Hu, H., Cooley, R., and Borowczak, M.}
\newblock Stealing machine learning parameters via side channel power attacks.
\newblock In {\em IEEE Computer Society Annual Symposium on VLSI\/} (2021),
  pp.~242--247.

\bibitem{xiang2020open}
{\sc Xiang, Y., Chen, Z., Chen, Z., Fang, Z., Hao, H., Chen, J., Liu, Y., Wu,
  Z., Xuan, Q., and Yang, X.}
\newblock Open {DNN} box by power side-channel attack.
\newblock {\em IEEE Transactions on Circuits and Systems II: Express Briefs\/}
  (2020), 2717--2721.

\bibitem{yan2020cache}
{\sc Yan, M., Fletcher, C.~W., and Torrellas, J.}
\newblock Cache telepathy: Leveraging shared resource attacks to learn {DNN}
  architectures.
\newblock In {\em Usenix Security\/} (2020), pp.~2003--2020.

\bibitem{yu2020deepem}
{\sc Yu, H., Ma, H., Yang, K., Zhao, Y., and Jin, Y.}
\newblock {DeepEM}: Deep neural networks model recovery through {EM}
  side-channel information leakage.
\newblock In {\em HOST\/} (2020), pp.~209--218.

\bibitem{zhang2016maximizing}
{\sc Zhang, H., and Hoffmann, H.}
\newblock Maximizing performance under a power cap: A comparison of hardware,
  software, and hybrid techniques.
\newblock In {\em ASPLOS\/} (2016), pp.~545--559.

\bibitem{zhang2014cross}
{\sc Zhang, Y., Juels, A., Reiter, M.~K., and Ristenpart, T.}
\newblock Cross-tenant side-channel attacks in paas clouds.
\newblock In {\em CCS\/} (2014), pp.~990--1003.

\bibitem{zhang2021stealing}
{\sc Zhang, Y., Yasaei, R., Chen, H., Li, Z., and Al~Faruque, M.~A.}
\newblock Stealing neural network structure through remote {FPGA} side-channel
  analysis.
\newblock {\em IEEE Transactions on Information Forensics and Security\/}
  (2021), 4377--4388.

\bibitem{zhang2021red}
{\sc Zhang, Z., Liang, S., Yao, F., and Gao, X.}
\newblock Red alert for power leakage: Exploiting intel rapl-induced side
  channels.
\newblock In {\em Proc. AsiaCCS\/} (2021), pp.~162--175.

\bibitem{zhu2021hermes}
{\sc Zhu, Y., Cheng, Y., Zhou, H., and Lu, Y.}
\newblock Hermes attack: Steal {DNN} models with lossless inference accuracy.
\newblock In {\em Usenix Security\/} (2021).

\end{thebibliography}

\appendices

\section{Domain Knowledge}\label{app:domain}
As shown in ~\autoref{tab:hyperparameters}, only three layer types have different layer-wise hyperparameters. 
In the following, we introduce all these layer-wise hyperparameter and their correlations based on the domain knowledge of popular and complex network families (e.g., MLP, AlexNet, VGG and ResNet), aligned with the SOTA~\cite{maia2021can}.

\mypara{A Convolutional Layer} 
This layer can have 7 hyperparameters, i.e., $C_{\rm in}$, $C_{\rm out}$, $K$, $S$, $P$, $D$, and $G$, which are introduced below:

\vspace{2pt}\noindent$\bullet$ $C_{\rm in}$ is the number of channels of an input. For the first layer, it is known as
1 or 3 dependent on whether an input image is grey or color. For other layers, $C_{\rm in}$ is the same as $C_{\rm out}$ of a preceding layer, indicating $C_{\rm in}$ can be decided when its preceding layer's $C_{\rm out}$ is known.

\vspace{2pt}\noindent$\bullet$ $C_{\rm out}$ is the channel number of an output feature map, equaling to the number of convolutional kernels. It is often selected from a wide range of discrete integer values.
Generally, $C_{\rm out}$ is set based on some heuristics. For example, $C_{\rm out}$ of popular networks such as VGG and ResNet is set to be $2^n$ where $n$ can be from $\{4,5, 6, \cdots\}$.

\vspace{2pt}\noindent$\bullet$ $K$ is the size of a convolutional kernel, which is often an odd number.
Generally, $K$ does not exceed 7~\cite{maia2021can}.

\vspace{2pt}\noindent$\bullet$ $S$ is the stride, indicating the amount of movement when a convolutional kernel moves.   
By default, it is set to 1 by default or 2 if the convolutional kernel is used to perform downsampling. 

\vspace{2pt}\noindent$\bullet$ $P$ is the size of padding, which is often added to the edges of feature maps to allow for more space for a convolutional kernel to cover the feature maps.
By default, $P = \lfloor (K - 1) / 2 \rfloor \times D$, with $D$ as the dilation.

\vspace{2pt}\noindent$\bullet$ $D$ is the space between kernel elements and $G$ is the number of blocked connections from the input channel to the output channel. 
By default, $D$ and $G$ are 1 for a classification task. 
Generally, $D\ne 1$ is used in semantic segmentation networks, and $G\ne 1$ is used in lightweight networks, e.g., tiny machine learning on edge devices. 
Thus, both $D$ and $G$ are 1 in our target model architectures.

\mypara{A MaxPooling Layer} 
This layer can have 4 hyperparameters, i.e., $K$, $S$, $P$ and $D$, which have similar meanings to that of the 
convolutional layer but different values. 
Specifically, $K$ is picked from $\{2,3\}$. $S$ is from $\{1,2\}$. $P$ is from $\{0,1\}$. $D$ is set to 1.

\mypara{A Linear Layer} 
This layer can have 2 hyperparameters, i.e., $F_{\rm in}$ and $F_{\rm out}$. 
Specifically, $F_{\rm in}$ is the input-feature size. Similar to $C_{\rm in}$, its value can be decided by $C_{\rm out}$ or $F_{\rm out}$. $F_{\rm out}$ is the output-feature size that is either equal to the number of classes in the last linear layer, or $2^n$ where $n$ can be $\{4,5,6,\cdots\}$ for other linear layers. 

\begin{figure}
    \centering
    \includegraphics[trim=0 0 0 0,clip,width=0.40\textwidth]{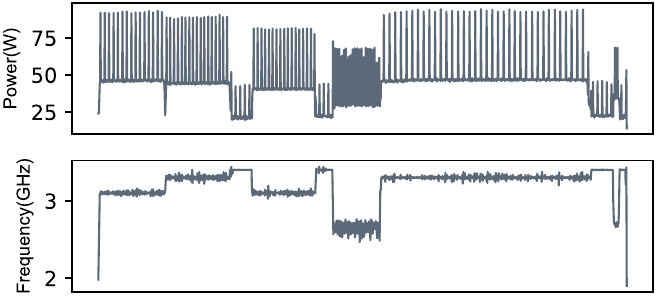}
    \caption{Time-series traces extracted from the RAPL-based power side channel (top)  and DVFS-enabled CPU frequency side channel (bottom), respectively. While they look different to each other, our proposed learning framework can learn each correlation between
the time-series traces and DNN model architectures in the offline phase and achieves high attack performance in the online phase.}
    \label{fig:power_freq}
    \vspace{+4mm}
\end{figure}

\end{document}